\documentclass[twocolumn]{aastex7}
\usepackage[skip=0.5ex]{subcaption}
\usepackage{amsmath,amstext}
\usepackage[T1]{fontenc}
\usepackage{apjfonts}
\usepackage[figure,figure*]{hypcap}
\usepackage{multirow}
\usepackage{subfiles}
\usepackage{graphicx}
\usepackage{longtable}
\usepackage{rotating}
\DeclareGraphicsExtensions{.png,.pdf}

\newcommand{\RomanNumeralCaps}[1]{\MakeUppercase{\romannumeral #1}}

\defcitealias{Bensby2014}{B14}
\defcitealias{Bedell2018}{B18}
\defcitealias{Battistini2016}{B16}

\begin{document}

\title{Four Elements to Rule Them All: Abundances are Rigidly Coupled in the Milky Way Disk}

\correspondingauthor{Jennifer Mead}
\email{jennifer.mead@columbia.edu}


\author[0009-0006-4744-2350]{Jennifer Mead}
\affiliation{Department of Astronomy, Columbia University, New York, NY 10027, USA}
\email{jennifer.mead@columbia.edu}

\author{Rebeca De La Garza}
\affiliation{Columbia College, Columbia University, New York, NY 10027, USA}
\email{rd3055@columbia.edu}

\author[0000-0001-5082-6693]{Melissa Ness}
\affiliation{Research School of Astronomy \& Astrophysics, Australian National University, Canberra ACT 2611, Australia}
\affiliation{Department of Astronomy, Columbia University, New York, NY 10027, USA}
\email{melissa.ness@anu.edu.au}

\keywords{Chemical Abundances; Chemical Tagging; Solar Neighborhood; Dimensionality Reduction, Milky Way Disk, Neutron Capture}

\begin{abstract}
Chemical tagging is a central pursuit of galactic archaeology, but requires sufficiently discriminative abundances to uniquely identify sites of star formation. This task is complicated by intrinsic scatter among conatal stars, inter-element correlations, imprecise abundance measurements, and systematics across stellar evolutionary states.  In this work, we formalize the abundance correlation structure of the disk by quantifying the amplitude of information in individual element abundances once a subset is known, and map inter-element residual correlations to uncover hidden signatures of nucleosynthesis. We use two datasets of 79 (593) stars across $-0.15<\rm [Fe/H]<0.13$ ($-1<\rm [Fe/H]<0.41$) with measurements of 30 (19) element abundances of solar neighborhood stars, including 11 (7) light and $\alpha$, 7 (3) Fe-peak, and 12 (9) neutron capture elements.  With a simple linear regression model, we predict most $\alpha$ and Fe-peak element abundances within 0.03~dex ($\sim7\%$), and neutron capture elements within 0.05~dex ($\sim10\%$). Including first and second peak \textit{s}-process elements as predictors improves most neutron capture element predictions to within 0.02~dex (5\%), although no predictive power is gained by including an \textit{r}-process element.  We uncover strong (anti-)correlations in small residual abundances between and within element families. Our finding that disk abundance space is rigidly coupled, from light to heavy elements, implies chemical tagging is infeasible at $>2\%$ precision for $\sim$30 elements. However, the residual structure encodes fingerprints of star formation history, inherited from nucleosynthesis and environmental variations, and provides critical constraints for chemical evolution models. Future disk surveys must achieve sub-2-5\% precision in 30+ elements to access this independent information.
\end{abstract}

\section{Introduction}
As the Universe has evolved from the first, metal-free, stars hundreds of millions of years after the Big Bang to the new stars we witness forming in our own Galaxy, each generation of stars has added their own contribution to the chemical composition of all subsequent generations of stars.  Elements have different nucleosynthetic origins, and different nucleosynthetic channels produce different families of elements. Each element is typically produced by more than one source and the relative source contribution is a function of time \citep{Kobayashi2020}.

Core-collapse supernovae (CCSNe) primarily eject $\alpha$-elements [e.g. magnesium (Mg), silicon (Si), calcium (Ca), titanium (Ti)], which are produced in the cores of massive stars ($> 8\,M_{\odot}$) \citep{Timmes1995,Kobayashi2006}. On the other hand, Fe-peak elements [e.g. manganese (Mn), chromium (Cr), iron (Fe), nickle (Ni)] are primarily produced by Type Ia supernovae (SNIa) \citep{Kobayashi2009}. However, while elements are grouped into nucleosynthetic families, and some, like Mg, are considered to be `pure' CCSNe elements, most elements, including Fe, Si, Ca, Ti, Mn, Cr and Ni have contributions from both SNIa and CCSNe \cite{Johnson2019,Kobayashi2020}.

Neutron capture elements have a variety of nucleosynthetic sources: \textit{s}-process elements [e.g. strontium (Sr), yttrium (Y), barium (Ba), lanthanum (La)] are produced during the AGB phase of stellar evolution for low mass stars ($<8\,M_{\odot}$) \citep{Busso1999,Herwig2005,Karakas2014}, and \textit{r}-process elements (e.g. europium (Eu), gadolinium (Gd), dysprosium (Dy)) are proposed to originate from binary neutron star mergers \citep{Lattimer1974,Rosswog1999,Goriely2011,Wanajo2014}, though other sites of formation have been proposed, including neutrino driven winds from core-collapse supernovae and other exotic sources such as electron-capture supernovae, magneto-rotational supernovae with jets, collapsars, and hypernovae \citep[][and references therein]{Cowan2021rprocReview}. While the neutron capture elements may be dominated by one pathway, most have partial r- and \textit{s}-process contributions.

Laying out the chemical history of the Universe is of course not as simple as measuring element abundances and mapping them to ratios of different nucleosynthetic channels due to multiple nucleosynthetic pathways for all elements \citep{Kobayashi2020}, and the potential metallicity- and mass-dependence of stellar yields \citep{WoosleyWeaver1995}.  While mass- and metallicity-dependent yields only have a weak effect on galaxy scale properties (e.g. star formation rate, total metallicity, stellar masses, etc.), these variations in yields have a stronger influence on the spread of individual abundances, particularly at low metallicity \citep{Muley2021}.

Comparisons between measured element abundances and chemical evolution models -- which track the production, distribution and enrichment of a galaxy over time using yield prescriptions from different sources -- offer a route to try to use measured abundances to learn the star formation history of the Milky Way \citep[e.g.][]{Chiappini2002, Molla2019, Kobayashi2020, Johnson2021, Matt2021, Buck2025}. The star formation history is typically parameterized by variables including the initial mass function, star formation efficiency, and star formation rate. These models are powerful tools than can successfully explain a number of global empirical trends and identify drivers of abundance signatures. However, there are substantial uncertainties and necessary assumptions made in the parameterization of star formation, element enrichment, and in stellar yields. Stellar yields alone have uncertainties relating to mass and metallicity dependencies, rotation, magnetic fields and mass loss. Additionally, the neutron flux densities that strongly influence \textit{s}-process element production are typically highly uncertain, and details of the stellar explosions can strongly influence element production. Therefore, in detail, there can be substantial discrepancies between chemical evolution models and the ensemble of empirical measurements of populations of stars. Even individual stars cannot be reproduced by yields from chemical evolution models; the abundance pattern of the Sun cannot be matched with theoretical models, nor can the ratio of two alpha elements [Mg/Si] in typical stars in the disk \citep{Chempy, Blancato2019, Weinberg2024}. Therefore, while global fits between data and models show discriminating power between models, given the uncertainties and assumptions required, it is not surprising there are discrepancies between observed element abundances and those generated in galactic chemical evolution models.

Currently lacking in the theoretical landscape are models that track the joint production and distribution of a large number of elements that include constraints based on their measured distributions and correlations. Other approaches include the use of simulations to trace global abundance trends although these typically look to make overall comparisons that look to interpret results rather than directly and accurately reproducing the detailed abundance measurements \citep[e.g.][]{Buck2021,AC2023, BB2024}.

In the present large-data era, whereby hundreds of millions of measured abundances are available, the data offer strong overall constraints for galactic chemical evolution models.  In addition, survey  data offer the means to test the premise of chemical tagging, i.e. to examine the dimensionality of the large vector of individual abundances. Recent work has demonstrated that the abundance space of the Milky Way disk is far more correlated than anticipated and that there are signals of detailed nucleosynthesis captured in the abundance measurements.  

In one of the first formalized data-driven approaches to nucleosynthesis, \citet{Casey2019} introduced a model where just six latent factors, four of which were consistent with known nucleosynthetic processes, were able to cluster stars in this space. \citet{Weinberg2019,Weinberg2022} introduced a 2-process model that was able to describe APOGEE \citep{apogee_overview} DR17 abundances as the sum of a core-CCSNe component and a SNIa component.  Examining the residuals of their model, they find two correlated element groups, and work from \citet{Ting2022} suggests that one must condition on at least seven elements from APOGEE before residual correlations are reduced to the level of observational uncertainties.  Using GALAH+ DR3 \citep{GALAH,GALAHDR3}, \citet{Griffith2022} found that a three-process model that includes a component for asymptotic giant branch (AGB) nucleosynthesis was better able to fit Ba and Y abundances, and that abundances could be predicted from a model to within 0.02-0.05~dex ($5-12\%$). Using APOGEE measurements of element abundances and stellar parameters alone, \citet{Ness2022} employed a simple linear regression to model 8 elements generated in SNIa and CCSNe (Si, O, Ca, Ti, Ni, Al, Mn, Cr) and learned they can be predicted to within $\sim$0.02 dex (or 5\%) using only 2 elements (Fe, Mg), for stars across the entire disk. They report only marginally higher predictive precision within birth clusters in these elements and connect this residual amplitude of 5\% to the level of homogeneity of the star-forming disk in supernovae elements, at fixed birth radius and age. \citet{Griffith2024} similarly find that the abundance space of the disk in 20 elements measured by APOGEE can be described with a 2-process model to within 0.02-0.05 dex (5-12\%). They further demonstrate that small amplitude residuals in elements such as Na, C+N, Ni, Mn, and Ce show correlations with different disk populations in the Milky Way. This is indicative that the residuals away from models employing a small number of elements or sources capture differences in the star formation history that link to structures that have been identified.

In this work, we quantify the dimensionality of the element abundance space of the Milky Way disk. For this pursuit, high-precision, high-fidelity measurements are key. Large surveys have transformed our understanding of the Milky Way, and have been a driving force behind substantial progress in data processing and achieving more accurate and precise abundance measurements. By exceeding their initially ambitious goals of measuring individual elements to 0.05-0.1 dex (12-25\%) precision \citep[e.g.][]{GP2016},  subsets of the large survey data with the highest precision have revealed that the disk abundances appear to sit on a low-dimensional manifold, at least in the light, $\alpha$, Fe-peak elements. We use a complement of high-fidelity data to extend this analysis across the periodic table, where precision is key rather than large numbers of stars. The results presented in this paper therefore substantially expand on previous work by predicting abundances for higher-resolution, higher-fidelity data and for a larger set of elements. 

We use a small a sample of high-fidelity data (79 total stars) and a simple regression model to predict 30 element abundances from a subset of two to four. Such high-fidelity data with precision abundances (1-10\%) is a necessity for this task given the prior analyses that have revealed the small sub-10\% residual amplitude of abundances beyond a 2-source or 2-element model.  Additionally, we use a sample of 593 stars with abundance precisions $>20\%$ for a more restricted analysis. We go beyond the 20 elements examined previously and study the amplitude of intrinsic information in 30 elements once a subset are known, which we term \textit{conditional abundance residuals}. Furthermore,  we study the behavior of many (12) neutron capture elements, forged in both the \textit{r}- and \textit{s}-process,  and their impact as additional predictors in our models and as elements being predicted. Finally, we examine the correlation structure between conditional abundance residuals. These provide an effective readout of stellar nucleosynthesis, and are a powerful measurement to pinpoint element origins and variations in star formation over the disk, and to constrain theoretical models.

In Section \ref{sec:data}, we review the three sets of data that we use in our modeling: \citet{Bensby2014} (hereafter \citetalias{Bensby2014}), \citet{Battistini2016} (hereafter \citetalias{Battistini2016}; which we combine with \citetalias{Bensby2014}) and \citet{Bedell2018} (hereafter \citetalias{Bedell2018}).  We describe our model whereby we predict element abundances using a subset, and our calculations for the precision of the predictions of our model in Section \ref{sec:methods}. This is followed by an analysis of the precision, which we quantify with intrinsic scatter, for various combinations of stellar parameters in Sections \ref{sec:2-source} and \ref{sec:3-source}. We present a complete model to describe our data sample in Section \ref{sec:Full_source}, and discuss the implications of our results for understanding the origin of elements, and for the future of large scale efforts in galactic chemical evolution modeling and chemical tagging in Section \ref{sec:disc}. We conclude in Section \ref{sec:conc}.

\begin{figure*}
    \centering
    \includegraphics[width=\linewidth,trim={2cm 2cm 0cm 3cm},clip]{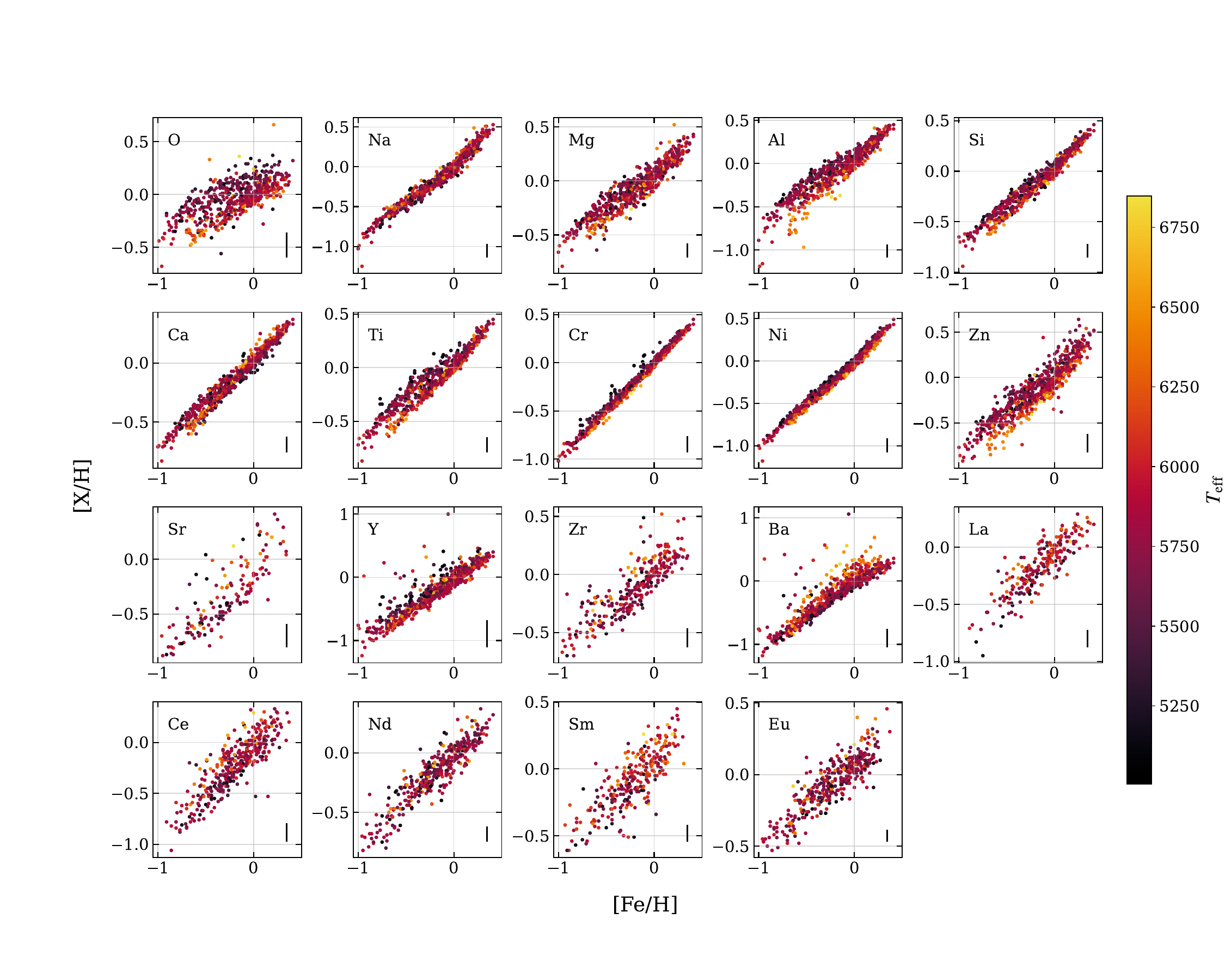}
    \caption{[X/H] vs. [Fe/H] for elements in \citetalias{Bensby2014} and \citetalias{Battistini2016}, colored by $T_{\rm eff}$.  The average error bars on [X/H] are show in the lower right of each plot.}
    \label{fig:Bensby_sum}
\end{figure*}

\section{Data} \label{sec:data}
We use data from two different samples: \citetalias{Bensby2014} with neutron capture abundances for this sample from \citetalias{Battistini2016}, and \citetalias{Bedell2018} to explore the predictive power of our models across different ranges in metallicity.  Both datasets consist of high-resolution spectra of main sequence stars in the solar-neighborhood obtained with different observational setups and criteria.

\subsection{\citet{Bensby2014} Data}
This sample from \citetalias{Bensby2014} is composed of 714 F and G dwarf stars in the solar neighborhood kinematically spanning the thin and thick disks, the stellar halo, and halo substructures. Spectra were taken by a series of surveys using various high resolution spectrographs ($R$ = 40,000 - 110,000), including FEROS, HARPS, UVES, SOFIN, FIES, and MIKE, with wavelength coverage within 3000-9200{\AA}. Detailed elemental abundances were obtained for 13 elements: O (oxygen), Na (sodium), Mg, Al (aluminum), Si, Ca, Ti, Cr, Fe, Ni, Zn (zinc), Y, and Ba using equivalent widths and parallel-plane atmospheres under the assumption of local thermodynamic equilibrium.  The stars in the sample have the following stellar properties:
\begin{itemize}
    \item $-2.6 <[$Fe/H$]< 0.5$ dex ($\overline{x} \approx -0.3$ dex),
    \item $4810< T_{\rm eff} <6960$ K ($\overline{x} \approx 5780$ K),
    \item $2.7<$ log$g$ $<4.76$ ($\overline{x} \approx 4.2$),
    \item $150 \le S/N \le 300$.
\end{itemize}
Figure \ref{fig:Bensby_sum} shows [X/H] vs. [Fe/H] for the \citetalias{Bensby2014} data, colored by T$\rm{_{eff}}$ for stars in this sample that are cross-matched with \citetalias{Battistini2016} in Section \ref{sec:Battistini}.

\subsection{\citet{Battistini2016} Data} \label{sec:Battistini}
The \citetalias{Battistini2016} data uses a subset of 593 stars from \citet{Bensby2014}, taken with the FEROS and MIKE spectrographs, to derive element abundances for seven elements: Sr, Zr (zirconium), La, Ce (cerium), Nd (neodymium), Sm (samarium), and Eu.  Spectra in this sample cover a wavelength range from 3500-9000\AA, with spectral resolutions of $R$ = 48,800 to 65,000, and $S/N>200$, with metallicities $-1<[\rm Fe/H]<0.41$.  Abundances [X/H] vs. [Fe/H] are plotted with the \citet{Bensby2014} data in Figure \ref{fig:Bensby_sum}.  For the remainder of this work, we used the cross-matched \citetalias{Bensby2014} and \citetalias{Battistini2016} dataset which has a total of 593 stars, and any reference to either the \citetalias{Bensby2014} or \citetalias{Battistini2016} alone is referencing this cross-match.

\begin{figure*}
    \centering
    \includegraphics[width=\linewidth,trim={2.5cm 2.5cm 0cm 3cm},clip]{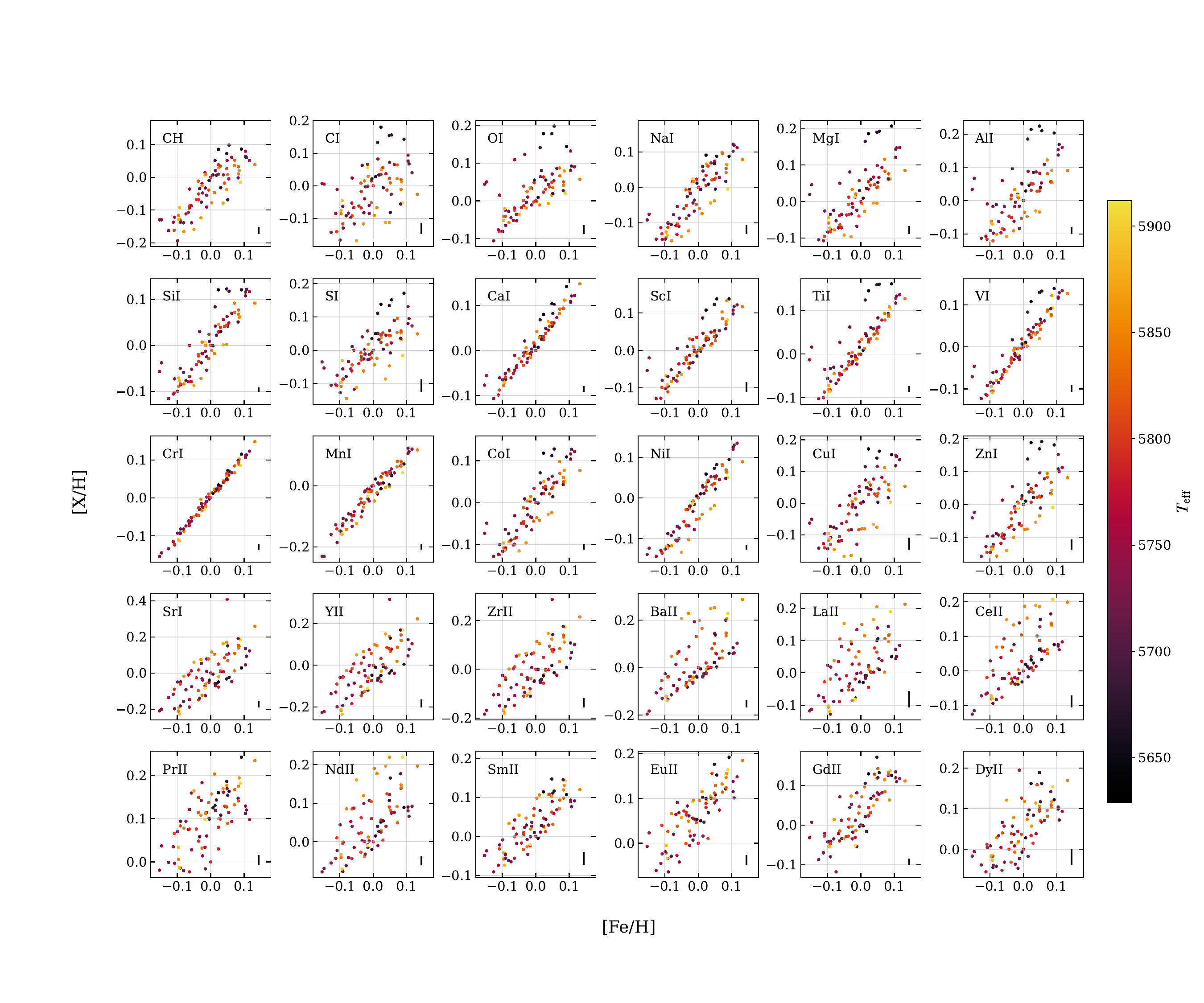}
    \caption{[X/H] vs. [Fe/H] for elements in \citetalias{Bedell2018}, colored by $T_{\rm eff}$.  5x the average error bar is shown in the lower right of each plot.}
    \label{fig:Bedell_summary}
\end{figure*}

\subsection{\citet{Bedell2018} Data}
The \citetalias{Bedell2018} sample is composed of 79 solar twins with spectra taken with the High Accuracy Radial velocity Planet Searcher (HARPS) spectrograph on the 3.6 meter telescope of the European Southern Observatory \citep[ESO;][]{Mayor2003}.  HARPS has a resolution of $R = 115,000$, covering wavelengths of $378-691$ nm.  The stars were chosen from a variety of sources for their similarity to the Sun, though some of the final derived parameters fell outside of these bounds:
\begin{itemize}
    \item $\lvert [\rm{Fe/H}]_* - [\rm{Fe/H}]_{\odot} \rvert \le 0.1$ dex ($\overline{x} \approx 0.0$ dex),
    \item $\lvert T_{\rm {eff,*}} - T_{\rm{eff,\odot}} \rvert \le 100$ K ($\overline{x} \approx 5780$ K)
    \item $\lvert \rm{log}$$g_* - \rm{log}$$g_\odot \rvert \le 0.1$ ($\overline{x} \approx 4.4$)
    \item $300 \le S/N \le 1800 \rm \, pix^{-1}$ ($\overline{x} \approx 800 \, \rm pix^{-1}$), achieved by stacking $\ge 50$ observations per star.
\end{itemize}

Elemental abundances and stellar parameters were obtained using a differential line-by-line equivalent width technique \citep[][]{Bedell2014}, with heavy element abundances from \citet{Spina2018}.  Abundances were measured for 30 independent elements: C \RomanNumeralCaps{1} (carbon), CH, O \RomanNumeralCaps{1}, Na \RomanNumeralCaps{1}, Mg \RomanNumeralCaps{1}, Al \RomanNumeralCaps{1}, Si \RomanNumeralCaps{1}, S \RomanNumeralCaps{1} (sulfur), Ca \RomanNumeralCaps{1}, Sc \RomanNumeralCaps{1} (scandium), Sc \RomanNumeralCaps{2}, Ti \RomanNumeralCaps{1}, Ti \RomanNumeralCaps{2}, V \RomanNumeralCaps{1} (vanadium), Cr \RomanNumeralCaps{1}, Cr \RomanNumeralCaps{2}, Mn \RomanNumeralCaps{1}, Fe, Co \RomanNumeralCaps{1} (cobalt), Ni \RomanNumeralCaps{1}, Cu \RomanNumeralCaps{1} (copper), Zn \RomanNumeralCaps{1}, Sr \RomanNumeralCaps{1}, Y \RomanNumeralCaps{2}, Zr \RomanNumeralCaps{2}, Ba \RomanNumeralCaps{2}, La \RomanNumeralCaps{2}, Ce II, Pr II (praseodymium), Nd \RomanNumeralCaps{2}, Sm \RomanNumeralCaps{2}, Eu \RomanNumeralCaps{2}, Gd \RomanNumeralCaps{2}, and Dy \RomanNumeralCaps{2}.  Figure \ref{fig:Bedell_summary} shows [X/H] vs. [Fe/H] for the \citetalias{Bedell2018} sample, colored by $T_{\rm eff}$.  For the remainder of this paper, we use only the neutral form of an abundance where both the neutral and ionized forms are available.

\subsection{Data Cuts}

We use the same data-cleaning procedure on both samples of data. To ensure the quality of the data, we remove all [X/Fe] measurements with uncertainties greater than 2 dex, resulting in a minimum average uncertainty of $0.08$~dex for Eu, maximum average uncertainty of $0.18$~dex for Y and overall average uncertainty of $0.11$~dex for the \citetalias{Bensby2014} and \citetalias{Battistini2016} combined data.  For the \citetalias{Bedell2018} data, this results in a minimum average uncertainty of $0.007$~dex for SiI, maximum average uncertainty of $0.026$~dex for LaII and overall average uncertainty of $0.013$~dex.  When making this cut, we do not remove all measurements for the star from our sample, rather we only remove the measurement for the relevant [X/H] model. Additionally, missing [X/H] measurements and values larger than 2 dex were removed, due to their unphysicallity.  On average from each [X/H], there were 15 measurements removed in \citetalias{Bensby2014}, 258 measurements in \citetalias{Battistini2016} and 0 measurements in \citetalias{Bedell2018}, preserving enough of each sample to model.

\section{Methods} \label{sec:methods}
The aim of this study is to (i) determine which stellar properties and abundances predict other measured elemental abundances, (ii) quantify to what precision elements are predicted from a subset, and (iii) to learn if the conditional abundance residuals of the $\rm (inferred \, abundance-measurement)$ are correlated, and if-so, what signatures these capture.

In this work, we opt to analyze abundances with respect to H. We note that for most elements, whether the abundances are themselves measured with respect to Fe or H does not change our results. However, measuring the abundances with respect to H removes the dependence on the variation in Fe production from different nucleosynthetic sources relative to other elements. 

\begin{figure*}
    \centering
    \includegraphics[width=\linewidth,trim={2cm 2cm 0cm 3cm},clip]{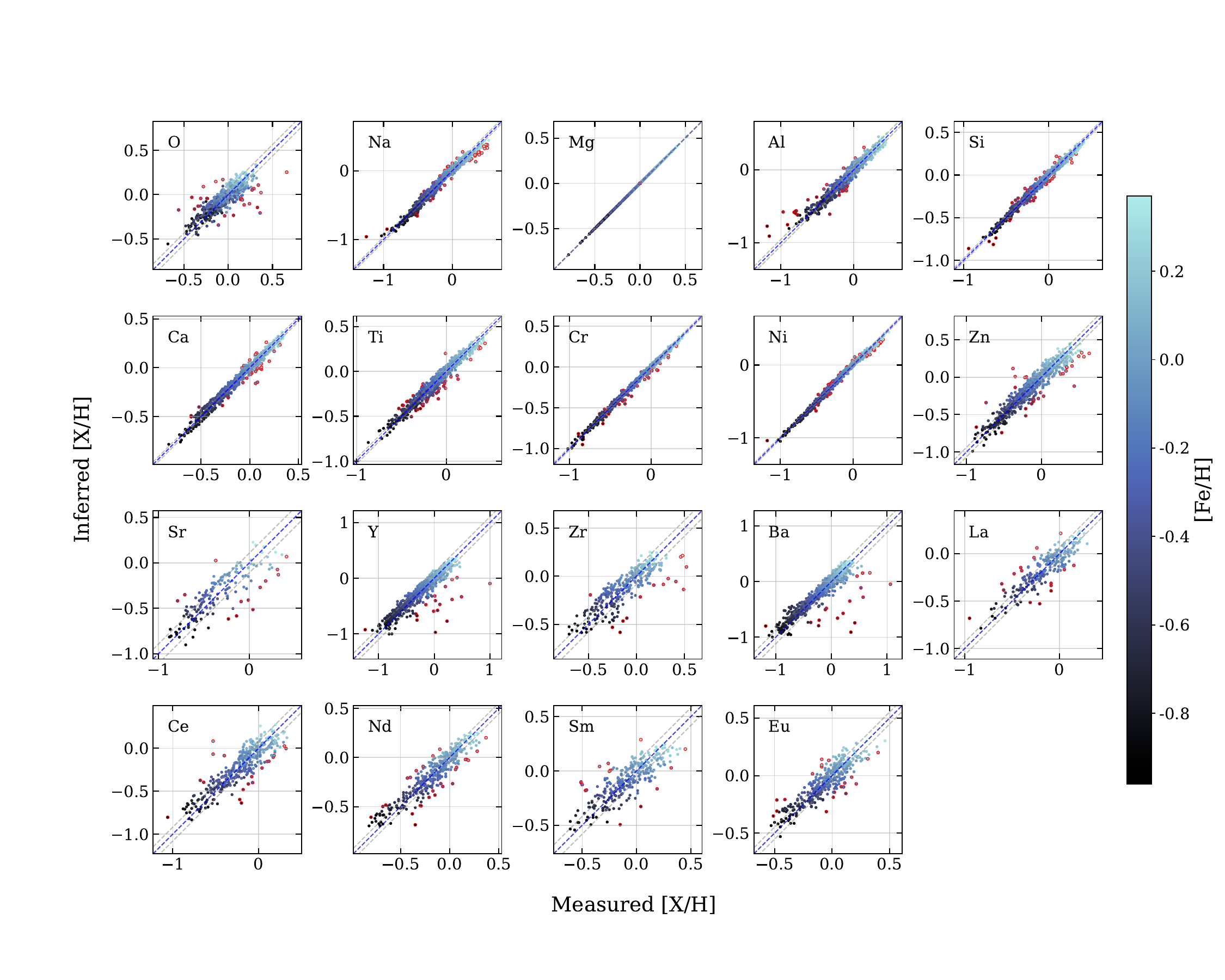}
    \caption{Fiducial model predictions on the \citetalias{Bensby2014} and \citetalias{Battistini2016} data, colored by [Fe/H]. The blue dashed line in each plot indicates the value for a model that perfectly predicts the data, with the gray dashed line denoting the $1\sigma$ spread in model residuals.  Inferred values with model residuals greater than $3\sigma$ are outlined in red.}
    \label{fig:Bensby_pred}
\end{figure*}

\subsection{Linear Regression} \label{sec:lin_regress}

We use linear regression to create multiple predictive models for all elements in our set.  The predictions generated from these models test the accuracy with which we can predict other abundances in our set. Our fiducial model (Equation \ref{eq:fid_model}) predicts abundances [X/H] using $T_{\rm eff}$, $\log g$, [Fe/H], [Mg/H]. 

\begin{equation} \label{eq:fid_model}
    m_{\rm ij} = a \,T_{\rm eff} + b \, {\rm log}g + c \, {\rm[Fe/H]} + d \, {\rm [Mg/H]} + g
\end{equation}

This fiducial model accounts for two known nucleosynthetic sources, CCSNe and SNIa (hence designating this a 2-source model), by including [Mg/H] and [Fe/H] respectively, and also accounts for systematics in abundances across evolutionary states by incorporating $T_{\rm eff}$ and log$g$. In order to prevent our predictive models from training on the data they are trying to predict, we generate a linear model, $m$, for each individual star, where model $m_{\rm ij}$ is the model for star $i$ that predicts abundance $j$ $(j: \rm [X/H])$, totaling $I \times J$ linear models for $I$ stars in the data set and $J$ measured elemental abundances.  Model $m_{\rm ij}$ is then generated by excluding the data from star $i$ from the linear regression and fitting the data for abundance $j$ from the remaining stars.  This ensures that the data from star $i$ are not used as part of the training set for model $m_{\rm ij}$. Each model $m_{\rm ij}$ is then used to predict abundance $j$ for star $i$.  The coefficients for each individual model do not change significantly from star to star, generally with standard deviations at least an order of magnitude lower than the mean, so the models provided in Table \ref{tab:models} of Appendix \ref{sec:app-model} average the coefficients for each individual star's model $m_{\rm ij}$ for a single model for the data, $m_j$, and we additionally provide standard deviations on each of the averaged coefficients.

\begin{figure*}
    \centering
    \includegraphics[width=\linewidth,trim={2.5cm 2.5cm 0cm 3cm},clip]{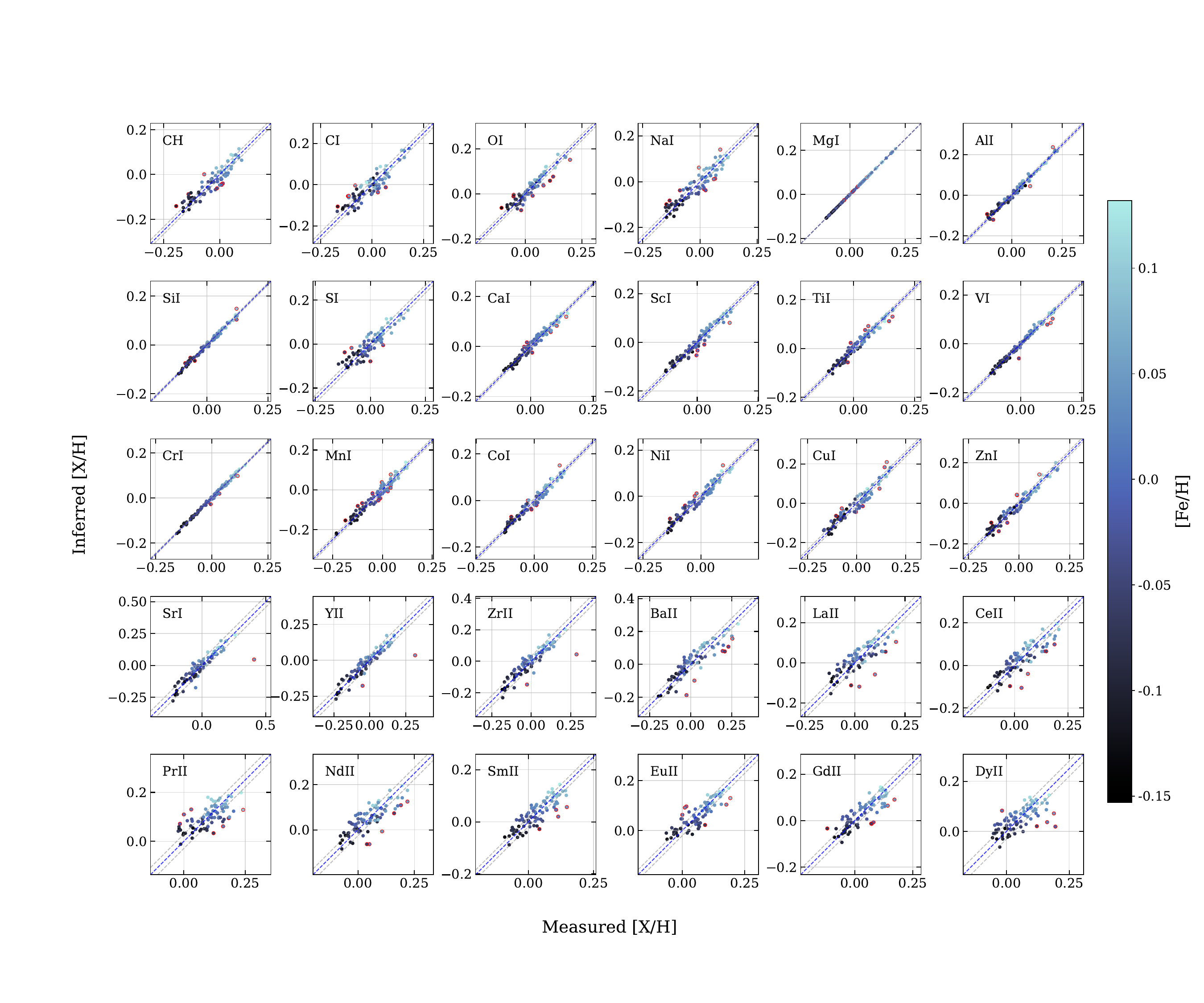}
    \caption{Fiducial model predictions on the \citetalias{Bedell2018} data, colored by [Fe/H]. The blue dashed line in each plot indicates the value for a model that perfectly predicts the data, with the gray dashed line denoting the $1\sigma$ spread in model residuals.  Inferred values with model residuals greater than $3\sigma$ are outlined in red.}
    \label{fig:Bedell_pred}
\end{figure*}

\subsection{Precision of the Model: Intrinsic Scatter}
To evaluate the prediction quality, or precision, of the model predictions, we calculate the intrinsic scatter, $s$, of model residuals. This intrinsic scatter is a measure of the dispersion of the predicted values around the true values after accounting for the contribution of measurement uncertainty. In order to account for the uncertainty on each individual abundance measurement, we maximize the joint probability distribution function (Equation \ref{eq:IS_mcmc}) by employing a Markov Chain Monte Carlo \citep[using \texttt{emcee};][]{emcee} to explore the posterior distribution on the intrinsic scatter. We calculate the intrinsic scatter using the residual between each data point and its respective model.

\begin{multline}
    \label{eq:IS_mcmc}
    P(x_{j,i}^o | \Bar{x_j},s_j,\delta x_{j,i})  = \\
    \prod_{i=1}^I 1/ \sqrt{2\pi(\delta x^2_{j,i} + s^2_j)} \cdot \rm{exp}(-\frac{(x^o_{j,i} - \Bar{x_j})^2}{2(\delta x^2_{j,i} + s^2_j)})
\end{multline}
Here, $dx_{j,i}$ is the uncertainty of each data point $i$ for abundance $j$, $s_j$ is the intrinsic scatter of the residuals about the model for abundance $j$, $x^o_{j,i}$ is the measured abundance $j$ of star $i$, and $\Bar{x_j}$ is the average measure of abundance $j$.

\section{2-Source Model} \label{sec:2-source}
\begin{figure*}
    \centering
    \includegraphics[width = \textwidth]{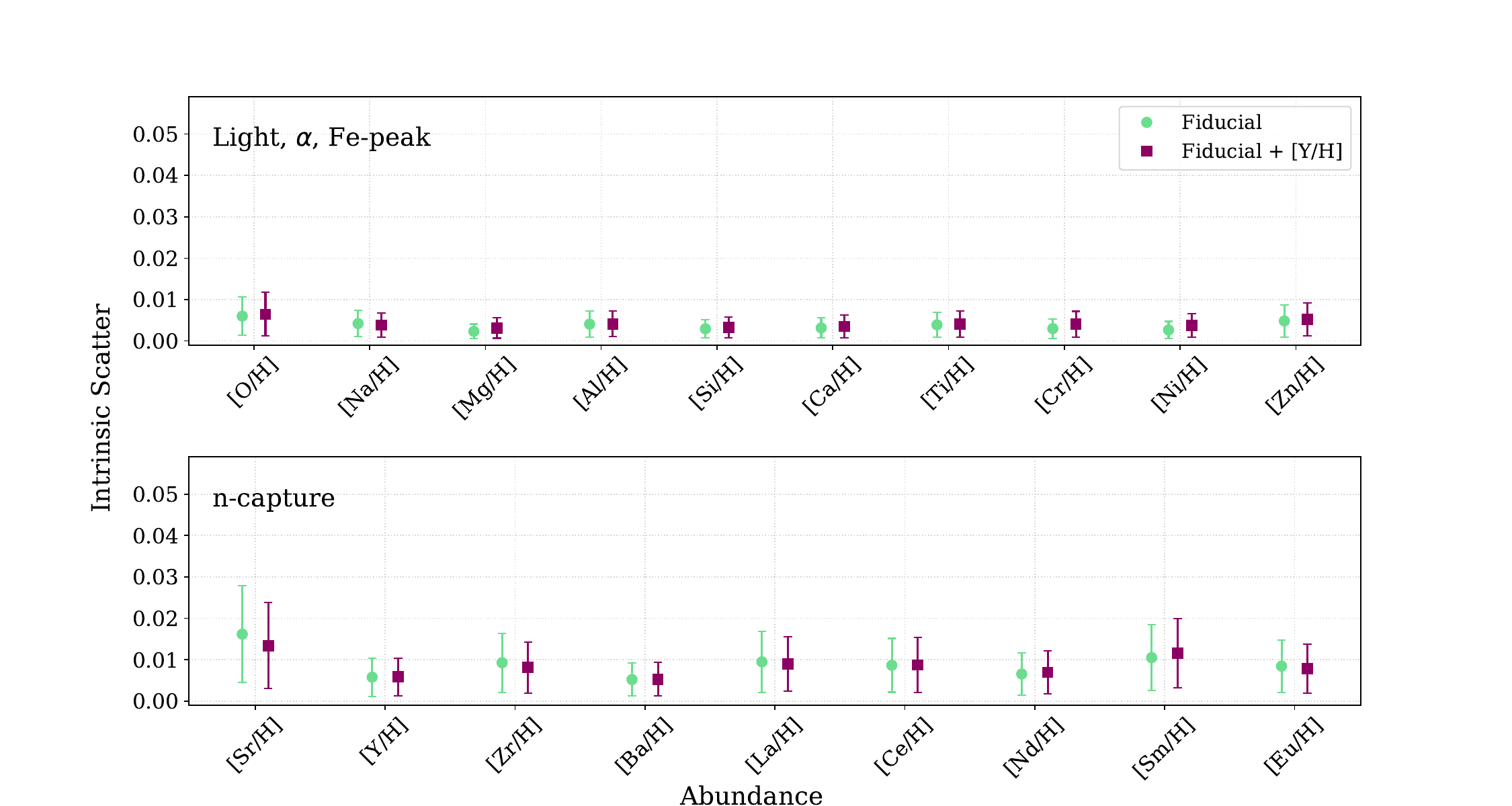}
    \caption{The intrinsic scatter of conditional abundance residuals between the inferred and measured data of \citetalias{Bensby2014}.  Shown are the intrinsic scatters for the fiducial 2-source model (green circles) and intrinsic scatters for a 3-source model that includes one representative first peak \textit{s}-process element, [Y/H] (purple squares).}
    \label{fig:IS_Bensby}
\end{figure*}

\subsection{Predictability of $\alpha$ and Fe-peak Abundances}
Using our fiducial model, we predict abundances [X/H] for all stars in the \citetalias{Bensby2014}/\citetalias{Battistini2016} and \citetalias{Bedell2018} datasets.  Figures \ref{fig:Bensby_pred} and \ref{fig:Bedell_pred} show each star's predicted and actual abundance measurement [X/H] for the \citetalias{Bensby2014}/\citetalias{Battistini2016} and \citetalias{Bedell2018} datasets respectively.  It is already clear from these figures that most elements in these datasets can be predicted well using our fiducial model\footnote{The apparent perfect prediction of Mg is a result of its inclusion in the model as opposed to any true predictive power.}. Although elements are well-predicted, there is visibly larger scatter in O, Zn, and all of the neutron capture elements in the \citetalias{Bensby2014} and \citetalias{Battistini2016} data, and in the \citetalias{Bedell2018} data for C, O, Na, S, and again many of the heavy elements, which are addressed in Section \ref{sec:3-source}.

\begin{figure*}
    \centering
    \includegraphics[width = \textwidth]{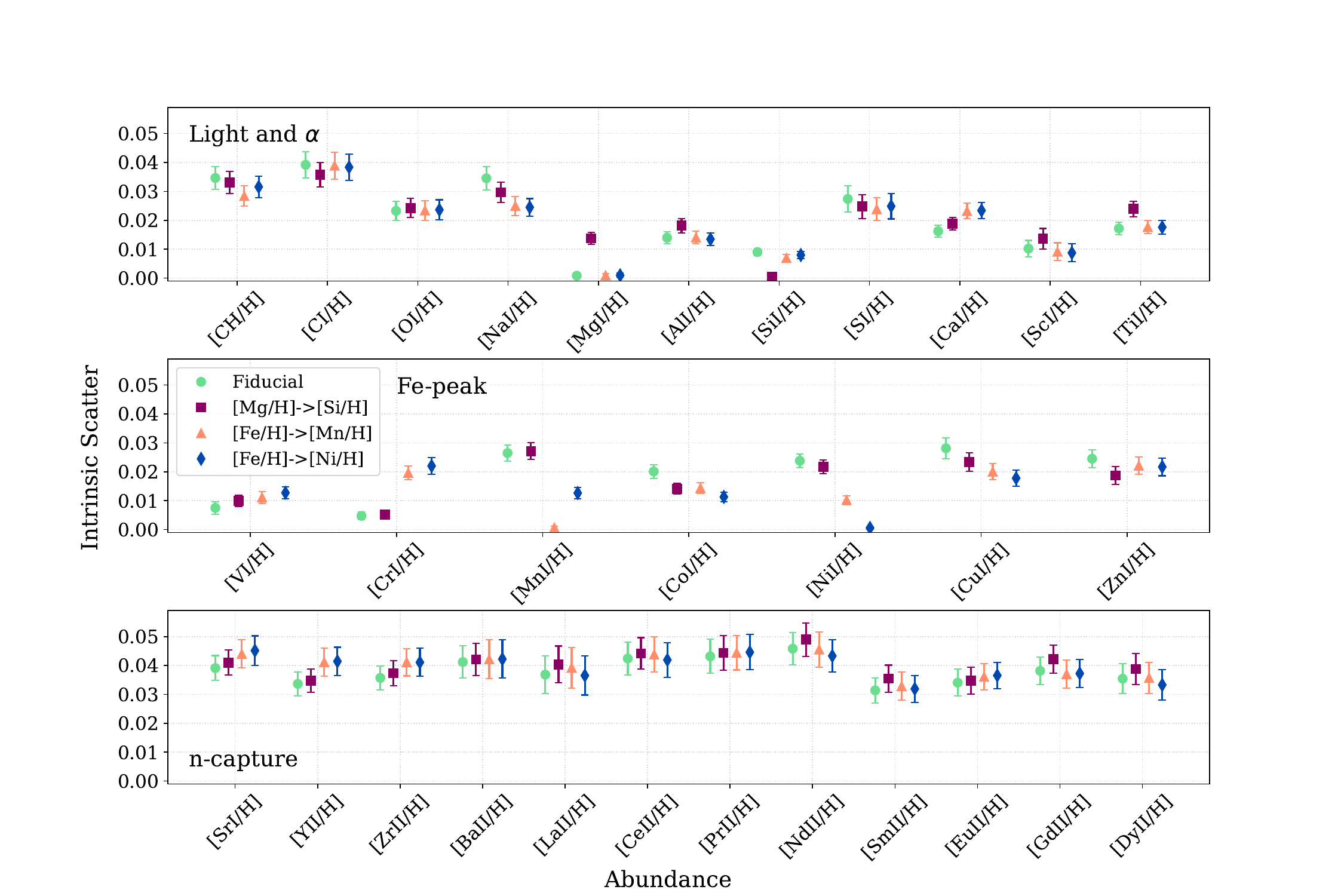}
    \caption{The intrinsic scatter of conditional abundance residuals between the inferred and measured data of \citetalias{Bedell2018}.  Shown are the intrinsic scatters for the fiducial model (green circles), as well as alterations on the fiducial model: substituting [Mg/H] for [Si/H] as a representative $\alpha$ element (purple squares), and substituting [Mn/H] (orange triangles) and [Ni/H] (blue diamonds) for [Fe/H] as a representative Fe-peak element.}
    \label{fig:IS_2proc}
\end{figure*}

Figures \ref{fig:IS_Bensby} and \ref{fig:IS_2proc} show the intrinsic scatter among the conditional abundance residuals on the predicted values of the fiducial 2-source model (green points) for \citetalias{Bensby2014}/\citetalias{Battistini2016} and \citetalias{Bedell2018}, respectively.  Figure \ref{fig:IS_2proc} reveals that we are able to predict most abundances from \citetalias{Bedell2018} within $\sim 0.02-0.03$~dex, or $5-7\%$, of the true value for the $\alpha$ and Fe-peak elements, with a maximum intrinsic scatter of 0.039~dex from [CI/H], a minimum of 0.005~dex from [CrI/H], and an overall average of 0.02~dex.  This is notably better than random dispersion, which we obtain by randomizing the inferred values, for which the minimum intrinsic scatter of the $\alpha$ and Fe-peak elements would be 0.06~dex with an overall average of 0.08~dex.

While it would be interesting to compare the accuracy in our predictions between the \citetalias{Bedell2018} data, which have a narrow metallicity range, and the \citetalias{Bensby2014} and \citetalias{Battistini2016} data, which have a wide metallicity range, the uncertainties in the \citetalias{Bensby2014} and \citetalias{Battistini2016} data dominate the residual scatter in the inferred values, driving the intrinsic scatter to misleadingly low values as seen in Figure \ref{fig:IS_Bensby} (recall that the average uncertainty of the \citetalias{Bensby2014} and \citetalias{Battistini2016} data is 0.11~dex, an order of magnitude larger than the calculated intrinsic scatters, compared to an average uncertainty of 0.013~dex in the \citetalias{Bedell2018} data which is similar to or lower than the typical intrinsic scatter by up to a factor of 5). For this reason, we primarily focus on the analysis of the \citetalias{Bedell2018} data from this point forward. However, it is worth noting that as is shown in Figures \ref{fig:Bensby_pred} and \ref{fig:Bedell_pred}, we see that both data samples are still well-predicted by our fiducial model and the \citetalias{Bensby2014} and \citetalias{Battistini2016} data have a minimum intrinsic scatter of 0.003~dex for [Ni/H], a maximum of 0.006~dex for [O/H] and average of 0.004~dex, where a randomized dispersion would have a minimum intrinsic scatter of 0.033~dex and average of 0.25~dex.  In particular, the ability to predict abundances using the fiducial model over the $-1\leq \rm[Fe/H]\leq0.41$ range of the \citetalias{Bensby2014} and \citetalias{Battistini2016} data reflects the common origins in nucleosynthetic sources, and the consistent patterns in enrichment over time.

\subsection{Outliers}
In Figures \ref{fig:Bensby_pred} and \ref{fig:Bedell_pred}, we indicate the $1\sigma$ cutoff on model residuals (gray dashed lines) and stars with model residuals greater than $3\sigma$ (red outlined points).  In the \citetalias{Bensby2014} and \citetalias{Battistini2016} data, we find that $70\%$ of stars that have model residuals that are $2\sigma$ outliers in one element abundance are outliers in at least one other element abundance, with $13\%$ of stars outliers in 5 or more element abundances.  At the $3\sigma$ cutoff, this drops to $40\%$ of stars being outliers in 2 or more element abundances, and only $2.5\%$ as outliers in 5 or more element abundances.  Two elements for which the outliers are of particular interest are Y and Ba, as outliers common to both could be a signature of mass transfer. We find that of the 23 stars that are $3\sigma$ outliers between Y and Ba, $50\%$ of these stars are outliers in both, but this is an insufficient sample to make conclusions about the frequency of mass transfer.

\subsubsection{Issue with Pr}
We note that from this point forward, we refrain from drawing conclusions about the element Pr in the \citetalias{Bedell2018} data.  As seen in Figure \ref{fig:Bedell_pred}, Pr demonstrates a tilt relative to the 1-to-1 line, overpredicting low abundances of [Pr/H], and underpredicting high abundances. As the errors for Pr are similar to those of other elements for this dataset, this is potentially indicative of a non-linear relationship between Pr and the stellar parameters used in the fiducial model.  We also note that \citet{Spina2018}, which reported the measurements we use for our analysis, also found significant scatter around the [X/Fe]-age relationship for Pr compared to other element abundances, suggesting that the production of Pr could be more closely tied to stellar birth site or is more sensitive to stochastic production such that it is not as closely correlated with stellar age. For these reasons, we exclude Pr from much of the remaining analysis due to uncertainty regarding the fidelity of its measurements.

\begin{figure*}[h]
    \centering
    \includegraphics[width=\linewidth,trim={7cm 4cm 6cm 6cm},clip]{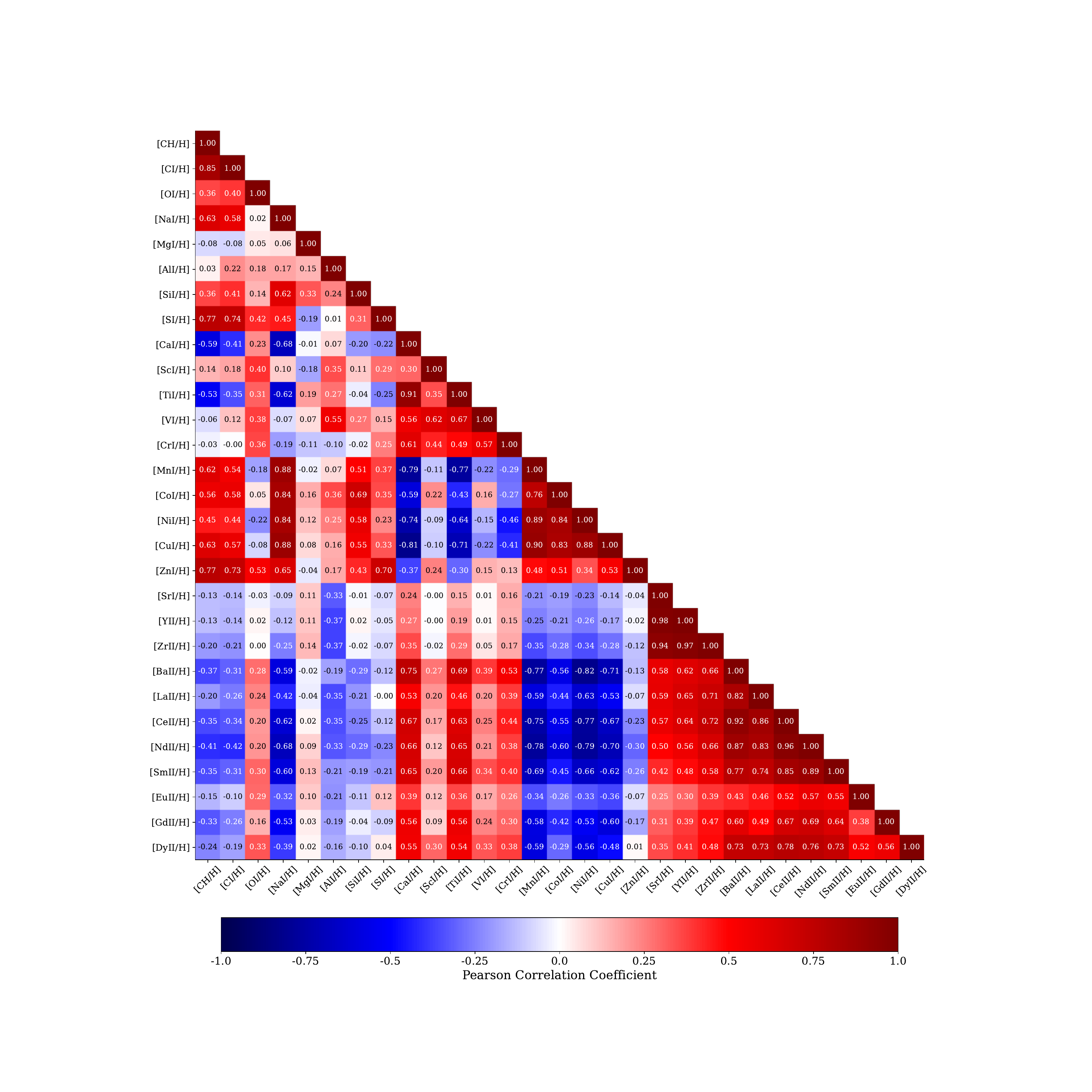}
    \caption{Pearson correlation coefficients of the [X/H] conditional abundance residuals from the fiducial model with every other [X/H] abundance residual. Red colors indicate positive correlation, blue colors indicate negative correlation, and darker shades indicate stronger (anti-)correlations.}
    \label{fig:res-res}
\end{figure*}

\subsection{Residual Correlations} \label{sec:res-corr}

In order to determine whether our models capture all the relevant inter-element correlations between elements, we look at the relationship between the conditional abundance residuals of each element abundance. The Pearson correlation coefficient of all conditional abundance residuals with each other are shown in Figure \ref{fig:res-res}, where red-shaded cells indicate a positive correlation, and blue-shaded cells indicate a negative correlation. It is immediately clear that the neutron capture elements have strong positive correlations among themselves.  These correlations mean that there is an additional parameter, unknown to the fiducial model, that would predict the data. We know from Figure \ref{fig:IS_2proc} that the neutron capture elements are significantly less-well predicted than the $\alpha$ and Fe-peak elements, with a minimum intrinsic scatter of 0.031~dex for [SmII/H], a maximum of 0.046~dex for [NdII/H] and an overall average of 0.038~dex. Combined, this indicates that we are missing a factor in our model that describes the neutron capture behavior. This suggests that the inclusion of one or more of the \textit{s}- and \textit{r}-process elements in our model would improve upon the predictions for the neutron capture elements.

Furthermore, many of the neutron capture elements, specifically the second \textit{s}-process peak (Ba, La, Ce, Nd, Sm), show correlations with the Fe-peak and some $\alpha$ elements. We find that the conditional abundance residuals for Ca, Sc, Ti, V, and Cr are positively correlated with the second peak \textit{s}-process elements, with Sc, V, and Cr showing weak ot moderate correlations, ranging from 0.12 (Sc-Nd) to 0.53 (Cr-Ba), and Ca and Ti showing moderate to strong correlations, ranging from 0.46 (Ti-La) to 0.75 (Ca-Ba).  Conversely, these same \textit{s}-process elements show moderate to strong anti-correlations with Fe-peak elements Mn, Co, Ni, and Cu ranging from -0.82 (Ni-Ba) to -0.44 (Co-La), and weak anti-correlations to no correlation with Zn.  Additionally, many of the Fe-peak elements (Mn, Co, Ni, Cu) are strongly positively correlated with each other, suggesting that Fe alone may be insufficient to capture the source that produces Fe-peak elements, potentially due to its mixed origins between CCSNe and SNIa.  While some of the correlation between elements could be attributed to noise, we find that the noise in the data would only shift the correlation coefficients by 0.08 on average, with the effect weakest for residuals with strong correlations (see Appendix \ref{sec:app-resnoise}).

Each of these residual correlations allude to information that is missing from our fiducial model.  This information may be other nucleosynthetic sources, as may be the case for the neutron capture elements, but these residuals may also encode information regarding the ages of stars and kinematic properties at birth as discussed in \citet{Griffith2024}.

\begin{figure*}[t!]
    \centering
    \includegraphics[width=\linewidth]{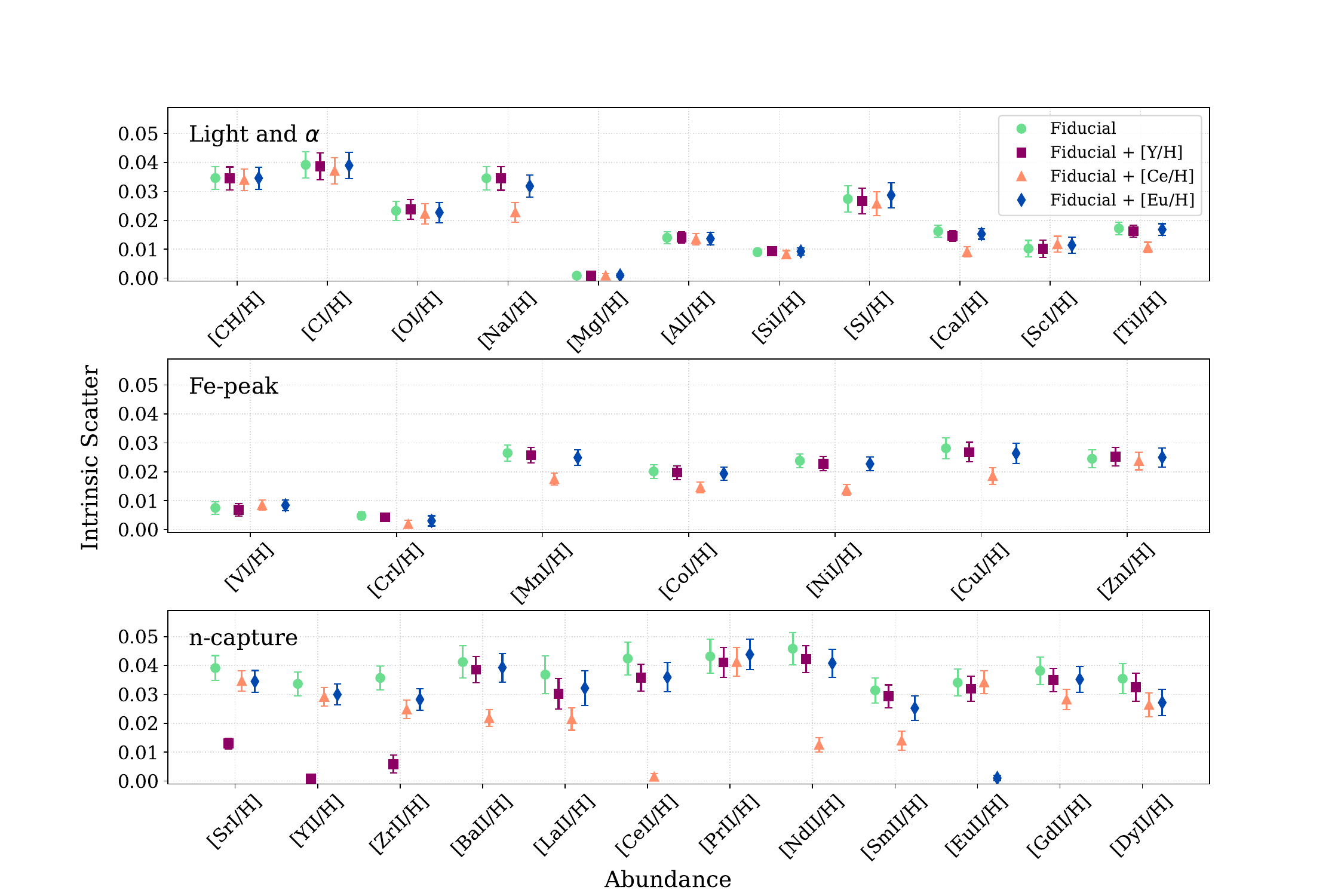}
    \caption{The intrinsic scatter of conditional abundance residuals between the inferred and measured data of \citetalias{Bedell2018}.  Shown are the intrinsic scatters for the fiducial model (green circles) as well as intrinsic scatters for 3-source models that demonstrate the impact on the intrinsic scatter of including a first peak \textit{s}-process element ([Y/H]; purple squares), a second-peak \textit{s}-process element ([Ce/H]; orange triangles), and an \textit{r}-process element ([Eu/H]; blue diamonds).}
    \label{fig:IS_3proc}
\end{figure*}

\subsection{Variations on the 2-source Model} \label{sec:2-proc_var}
Motivated by the inability of the fiducial 2-source model to fully explain the correlations between $\alpha$ and Fe-peak elements, we explore whether Mg and Fe are sufficiently representative elements of the nucleosynthetic channels that produce them.  We test the following modifications on the 2-source fiducial model:

\begin{equation} \label{eq:testFe_model1}
    m_{\rm ij} = a \,T_{\rm eff} + b \, {\rm log}g + c \, {\rm[Mn/H]} + d \, {\rm [Mg/H]} + g
\end{equation}
where [Fe/H] has been replaced with [Mn/H], an element more dominantly produced in SNIa \citep{Kobayashi2020},

\begin{equation} \label{eq:testFe_model2}
    m_{\rm ij} = a \,T_{\rm eff} + b \, {\rm log}g + c \, {\rm[Ni/H]} + d \, {\rm [Mg/H]} + g
\end{equation}
where [Fe/H] has been replaced with [Ni/H], an element with similar ratios produced from various nucleosynthetic sources as Fe \citep{Kobayashi2020}, and

\begin{equation} \label{eq:testalpha_model}
    m_{\rm ij} = a \,T_{\rm eff} + b \, {\rm log}g + c \, {\rm[Fe/H]} + d \, {\rm [Si/H]} + g
\end{equation}
where [Mg/H] has been replaced with [Si/H] to test the robustness of using Mg as a representative $\alpha$ element.

Figure \ref{fig:IS_2proc} shows the results of replacing the representative element for nucleosynthetic sources that produce $\alpha$ and Fe-peak elements.  For nearly all element abundances, substituting [Si/H] for [Mg/H] (purple squares) enables predictions to the same level of accuracy as our fiducial model, supporting that Mg and Si capture the same nucleosynthetic pathways, and that models could use them interchangeably.  However, substituting [Mn/H] (orange triangles) or [Ni/H] (blue diamonds) for [Fe/H] has a statistically significant effect on the Fe-peak elements, though they have little effect on the light, $\alpha$ and neutron capture elements.  Namely, the precision of the modified models is worse than the fiducial model for [V/H] and [Cr/H], but significantly better for [Mn/H], [Co/H], [Ni/H], and [Cu/H].  This improvement is perhaps expected given the strong residual correlations between these elements shown in Figure \ref{fig:res-res}. While we would not use these elements interchangeably with [Fe/H] as a representative Fe-peak element, the differences in the models potentially hint at further subtlety in the production pathways of these elements or strong connections to other birth properties that warrant future work in this area.

\begin{figure*}
    \centering
    \includegraphics[width=\linewidth,trim={1cm 2cm 1.9cm 5cm},clip]{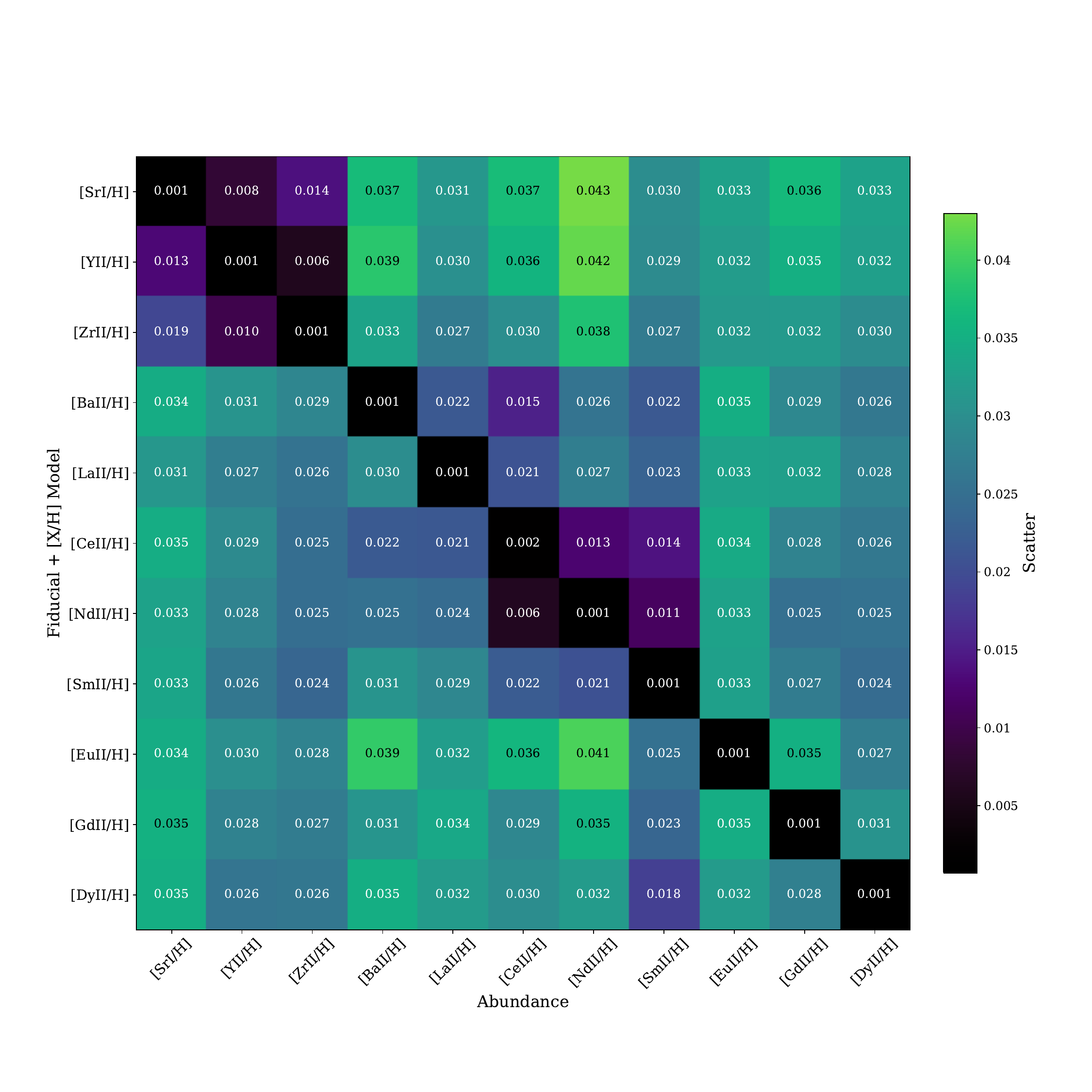}
    \caption{Intrinsic scatters on the inferred values for all neutron capture elements in \citetalias{Bedell2018}, for each 3-source model including a neutron capture abundance, [X$_{\rm sr}$/H].}
    \label{fig:bedell-s-proc}
\end{figure*}

\begin{figure*}
    \centering
    \includegraphics[width=\linewidth]{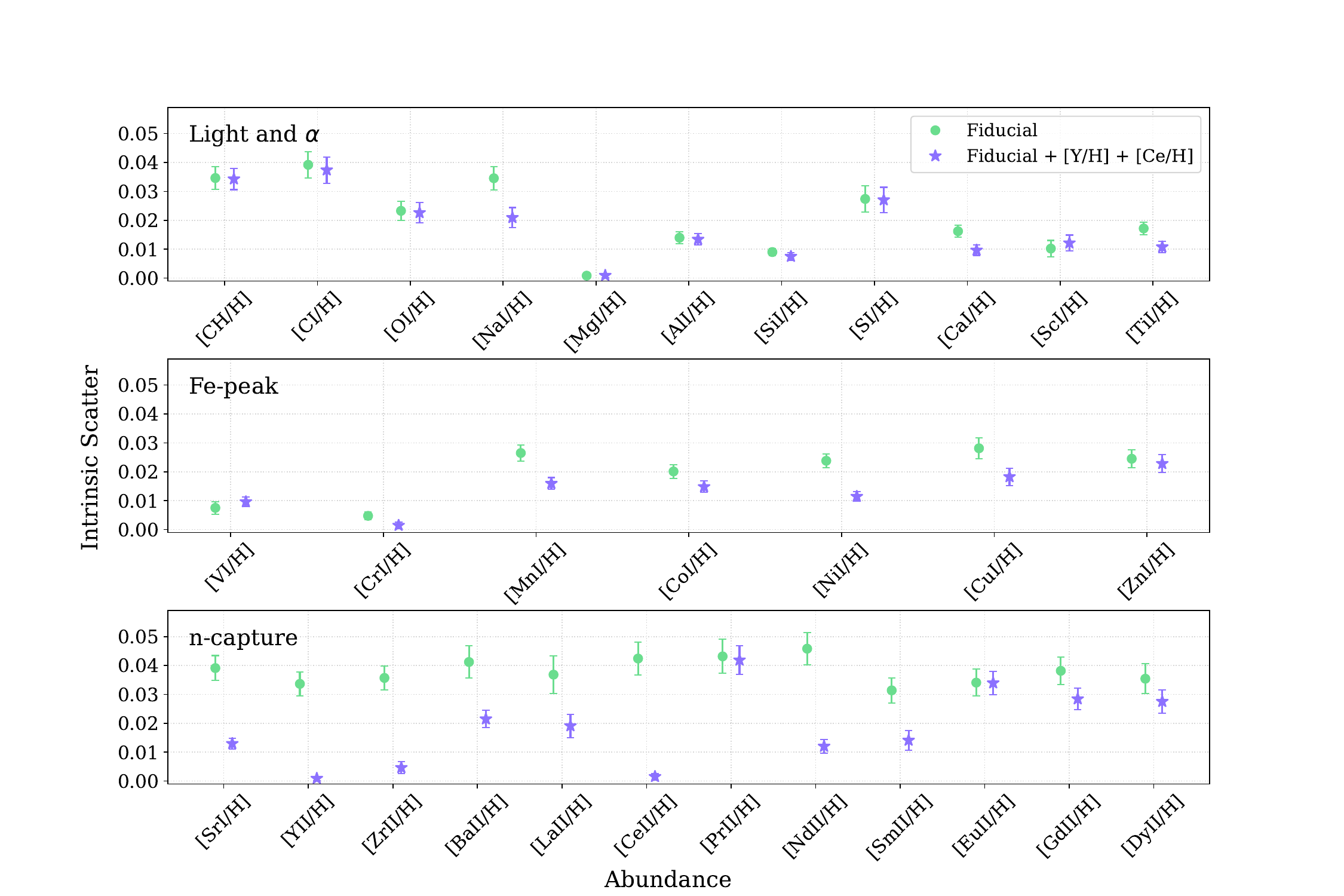}
    \caption{The intrinsic scatter of conditional abundance residuals between the inferred and measured data of \citetalias{Bedell2018}.  Shown are the intrinsic scatters for the fiducial model (green circles) as well as the intrinsic scatters for our full, 4-source model (purple stars) which includes, in addition to the fiducial model, contributions from a representative first peak \textit{s}-process element ([Y/H]) and a representative second peak \textit{s}-process element ([Ce/H]).}
    \label{fig:bedell-full}
\end{figure*}

\section{3-Source Model} \label{sec:3-source}
Informed by the correlations observed in the neutron capture elements in Figure \ref{fig:res-res}, we test additional linear models. Notably, sources of neutron capture elements are absent from the 2-source fiducial model that we use, and as such, their element abundances are not well predicted by that model. Therefore, we add an additional neutron capture element to our fiducial model, iterating through the elements. We test whether or not an additional source reduces the model intrinsic scatter as well as inter-element residual correlations, via the presumed accounting for additional nucleosynthetic channels.  We test the accounting for the \textit{s}-process and \textit{r}-process by including a neutron capture element in the model as follows:
\begin{equation} \label{eq:sproc_model}
    m_{\rm ij} = a \,T_{\rm eff} + b \, {\rm log}g + c \, {\rm[Fe/H]} + d \, {\rm [Mg/H]} + e \, {\rm [X_{sr}/H]} + g
\end{equation}
where $\rm{X_{sr}}$ represents any of the \textit{s}- or \textit{r}-process elements in the dataset,

In order to identify the element(s) that best represent the neutron capture elements, we substitute each neutron capture element in the dataset into $\rm [X_{sr}/H]$ in Equation \ref{eq:sproc_model}.  We test this for the \citetalias{Bensby2014} and \citetalias{Battistini2016} models, but find that as the scatter in the model residuals are primarily dominated by the uncertainties in the data, adding a neutron capture element cannot significantly improve the model. We show an example model using [Y/H] as a representative neutron capture element in Figure \ref{fig:IS_Bensby}.  Similarly, for the \citetalias{Bedell2018} data, we show intrinsic scatters for example models incorporating a first peak \textit{s}-process element ([Y/H]; purple squares), a second peak \textit{s}-process element ([Ce/H]; orange triangles), and an \textit{r}-process element ([Eu/H]; blue diamonds) in Figure \ref{fig:IS_3proc}. Figure \ref{fig:bedell-s-proc} shows the intrinsic scatter in the residuals for each model prediction for the \citetalias{Bedell2018} data, where the element abundance substituted into the model is on the \textit{y}-axis, and the element abundance predicted with each model is on the \textit{x}-axis.  Abundances are organized by atomic number.

\subsection{$s$--process} \label{sec:s-proc}
We first focus on the models which include \textit{s}-process element abundances (Sr, Y, Zr, Ba, La, Ce, Nd, Sm).  It is immediately clear that model--abundance pairs surrounding the diagonal in Figure \ref{fig:bedell-s-proc} have stronger predictive power, or lower intrinsic scatter in the residuals, than model-abundance pairs farther off the diagonal.  There initially appear to be three groups that have lower intrinsic scatter among themselves: (Sr, Y, Zr), (Ba, La, Ce), and (Ce, Nd, Sm). However, notably, the elements in the second \textit{s}-process peak (Ba, La, Ce, Nd, and Sm) predict each other fairly well, generally more than $1\sigma$ better than for other elements (e.g. [Ce/H] model shown in Figure \ref{fig:IS_3proc}).  Furthermore, models including a first peak \textit{s}-process element abundance (e.g. [Y/H] model in Figure \ref{fig:IS_3proc}) are capable of predicting other first peak \textit{s}-process element abundances more than $2\sigma$ better than the fiducial model, to an accuracy of about 2\%, versus 9\% with the fiducial model.  We see a similar improvement with the second peak \textit{s}-process elements, with at least a $2\sigma$ improvement over the fiducial model to an accuracy of $\sim2-5\%$ versus $\sim7-11\%$ in the fiducial model.

In Figure \ref{fig:IS_3proc}, we also show the effect of adding an \textit{s}-process element to the fiducial model on the predictive power for the $\alpha$ and Fe-peak elements, particularly those that showed strong residual correlations with the \textit{s}-process elements in the fiducial model.  While [Y/H], a representative first peak \textit{s}-process element does not improve the predictive power of the model for $\alpha$ and Fe-peak elements, [Ce/H], a representative second peak \textit{s}-process element, shows a statistically significant improvement, particularly for [Na/H], [Ca/H], [Ti/H], [Mn/H], [Co/H], [Ni/H], and [Cu/H], all of which were strongly correlated with the second peak \textit{s}-process elements in Figure \ref{fig:res-res}.

\subsection{$r$--process}
Unlike the \textit{s}-process elements, the inclusion of an \textit{r}-process element in the predictive model does not improve the predictive power for any element, including other \textit{r}-process elements, as shown in Figures \ref{fig:IS_3proc} and \ref{fig:bedell-s-proc}. We remain unable to predict \textit{r}-process elements better than $\sim0.03-0.04$~dex, within the statistical uncertainty of the fiducial model.  The lack of improvement in models accounting for the \textit{r}-process suggests that there may be no singular dominant source that produces \textit{r}-process elements.

\section{A Full 4-Source Model} \label{sec:Full_source}
Following the improved predictions from including first peak and second peak \textit{s}-process elements in the model, we present results from a full, 4-source, model:

\begin{multline}
    \label{eq:full_model}
    m_{\rm ij} = a \,T_{\rm eff} + b \, {\rm log}g + c \, {\rm[Fe/H]} + d \, {\rm [Mg/H]} \\
    + e \, {\rm [Y/H]} + f \, {\rm [Ce/H]} + g
\end{multline}
which includes an $\alpha$ ([Mg/H]), Fe-peak ([Fe/H]), first peak \textit{s}-process ([Y/H]), and second peak \textit{s}-process ([Ce/H]) element, representative of four different nucleosynthetic origins.  We predict all element abundances in \citetalias{Bedell2018}, and compare the accuracy of this full model to the fiducial model in Figure \ref{fig:bedell-full}.  The full model adopts all of the best predictions from the fiducial, and [Y/H] and [Ce/H] 3-source models, with the most notable improvements over the fiducial model occurring for the \textit{s}-process elements, and a subset of Fe-peak elements. Notably, the structure in the conditional abundance residual correlations among the \textit{s}-process elements, and between the \textit{s}-process elements and subset of $\alpha$ and Fe-peak elements has been reduced so most absolute correlations among this subset are below 0.3 (see Appendix \ref{sec:app-resfull}). With the exception of some light elements and the \textit{r}-process elements, the full 4-source model makes abundance predictions that are accurate to within $<5\%$.

\section{Discussion} \label{sec:disc}

\subsection{Mass and Metallicity Dependence of the \textit{s}-process}
In Section \ref{sec:s-proc}, our models are able to distinguish between two groups of elements: the first peak (light, \textit{ls}) and second peak (heavy, \textit{hs}) \textit{s}-process elements.  These two groups have physical origins which stem from the ratio of neutrons to seed Fe nuclei available to capture neutrons \citep[see reviews][and references therein]{Kappeler2011,Karakas2014,Lugaro2023}.  The neutron density is tied to the stellar mass and metallicity of the AGB star.  Low-mass AGB stars ($\sim 1-4 M_\odot$) produce neutrons primarily through the $^{13}{\rm C}(\alpha,n)^{16}{\rm O}$ reaction, which is active over long timescales ($\sim 10^4$ years) between thermal pulses, producing neutron densities $\sim10^7 \rm cm^{-3}$ and enabling the production of first peak \textit{s}-process elements, resulting in a higher [\textit{ls/hs}].  Stars $>3-4 M_\odot$ can reach temperatures exceeding $3\times10^8$K, triggering the $^{22}{\rm Ne}(\alpha,n)^{25}{\rm Mg}$ reaction which is only active for a few years following a thermal pulse, but produces higher neutron densities ($\sim 10^{13} \rm cm^{-3}$) that enable the production of second peak \textit{s}-process elements.  Higher-mass AGB stars are able to produce neutron densities high enough to achieve third peak \textit{s}-process elements such as lead (Pb).  Operating in conjunction with the stellar mass is the metallicity of the star, with stars of lower metallicities able to produce more \textit{hs} elements than their metal-rich counterparts due to the lower number of Fe seed nuclei and thus higher relative neutron-to-seed ratio.

The low neutron densities of low-mass AGB stars only enable \textit{s}-process neutron capture for isotopes of elements up to the fifth neutron magic number, $N=50$, where a neutron shell is completely filled in the nucleus leading to increased binding energy and stability \citep[][]{Kappeler2011,Karakas2014}.  This results in the first peak \textit{s}-process elements including Sr, Y, and Zr.  The higher neutron densities that occur in intermediate-mass AGB stars enable \textit{s}-process neutron capture elements to form up to next neutron magic number, $N=82$, resulting in the second peak \textit{s}-process elements, including Ba, La, Ce, Nd, and Sm.  The highest-mass (and low-metallicity) AGB stars have neutron densities high enough to also reach element isotopes with $N=126$, forming elements such as Pb and Bi (bismuth).

Our models have demonstrated that all of the \textit{s}-process elements at fixed (Fe, Mg) are correlated. Specifically, the most rigidly coupled \textit{s}-process elements are for those formed in environments with similar neutron density. For these, the Pearson correlation coefficient is up to 0.9 (see Figure \ref{fig:res-res}).  This means that at fixed enrichment from CCSNe and SNIa, the conditional residual abundance correlations capture the signatures of the distribution of mass and metallicities of the AGB stars enriching the ISM in the \textit{s}-process. We are subsequently able to use the first and second peak neutron capture sources as distinct channels of information in our model, which provides better predictive power than a single representative element to capture the \textit{s}-process.

\subsection{Time-scale Correlations between CCSNe and AGB Phase} \label{sec:timescales}

In Section \ref{sec:res-corr}, we identified strong (anti-)correlations between residuals of the \textit{s}-process elements, particularly the second peak, and $\alpha$ (Fe-peak) elements in our fiducial model, specifically strong positive correlations with elements that are produced in CCSNe with weaker correlations with elements that can be produced by massive star core nuclear fusion, and negative correlations with elements that are more dominantly produced by SNIa as classic Fe-peak elements.  By testing a 3-source model in Section \ref{sec:3-source}, which accounted for neutron capture elements, we found that those models that included a second peak \textit{s}-process element were able to improve the scatter in the residuals of the $\alpha$ and Fe-peak elements whose residuals were correlated with second peak elements by 0.5-1.5~dex.  The correlation between the $\alpha$ and Fe-peak elements with the second peak \textit{s}-process elements, and more intriguingly \textit{not} with the first peak \textit{s}-process elements, is an indicator that the information we are capturing with this correlation is, as expected, not related strictly to nucleosynthetic origin.  Additionally, the divide between the elements whose residuals that are positively and negatively correlated with the second peak \textit{s}-process elements suggests something similar.

We posit that we are in fact witnessing an overlap of timescales between massive--intermediate mass AGB stars and CCSNe.  Stars with masses $\geq 7M_\odot$, which primarily produce second and third peak \textit{s}-process elements, reach their AGB phase around 50 Myr \citep[][]{PARSEC2a,PARSEC2b}, which overlaps with the timescale for CCSNe for the lowest mass stars which undergo one.  In particular, we expect that the $\alpha$ and Fe-peak elements produced in core-burning in massive stars and in CCSNe are positively correlated with second-peak \textit{s}-process elements due to the overlapping timescales, and that Fe-peak elements produced in SNIa are anti-correlated with second peak \textit{s}-process due to the much longer timescale of SNIa \citep[500 Myr to 10 Gyr; e.g.][]{Ruiter2009,Maoz2010,Maoz2012}.

\subsection{Implications for Sites of \textit{r}-Process}

The inability to improve the predictive power of our model by including an \textit{r}-process element suggests (i) there is not one dominant source that produces a consistent pattern of \textit{r}-process abundances, and (ii) the \textit{r}-process elements are uncorrelated among and/or between sites of \textit{r}-process nucleosynthesis.  However, we note that this data sample only contains measurements for three \textit{r}-process elements, and that to identify correlations between \textit{r}-process, a sample with more \textit{r}-process elements would be required, particularly to identify any correlations that may exist between elements that may be dominantly produced in different sites of \textit{r}-process.

If it is true that the \textit{r}-process elements are uncorrelated or that only small groups of elements are correlated, these relationships could be used to constrain sites of \textit{r}-process for different elements, potentially by using timescales, as discussed in Section \ref{sec:timescales}.  In our fiducial model, Figure \ref{fig:res-res} shows that the \textit{r}-process is correlated in the same way as the \textit{s}-process to the $\alpha$ and Fe-peak elements, potentially suggesting that a dominant early source of \textit{r}-process occurs on timescales similar to CCSNe, such as collapsars \citep{Fujimoto2007,Siegel2019}, or magnetorotational SNe \citep{Winteler2012,Nishimura2017}, and disfavor neutron star mergers which occur on timescales of 100 Myr or longer \citep{Siegel2022,Maoz2025}.  

\subsection{Correlations with Other Stellar and Galactic Properties}

The strong inter-element conditional abundance residual correlations that we see at fixed conditions of Fe and Mg in Figure \ref{fig:res-res}, mean that the scatter seen in Figures \ref{fig:Bensby_pred} and \ref{fig:Bedell_pred} is not random, especially so for the neutron capture elements. These correlations must mean that the chemical enrichment history of stars in the disk is governed by a very small number of underlying parameters. These parameters obviously connect to element abundances and nucleosynthetic source, but are possibly best parameterized as neither. This rigid coupling of abundances further suggests that the star formation and enrichment of the Milky Way disk took place over constrained conditions, which may be encoded in the conditional abundance residuals.

It has been shown in previous work that abundances themselves along with the residuals of 2-process models are correlated with age, $Z_{\rm max}$, guiding radius, angular momentum, and vertical action \citep{Spina2018,Ness2022,Griffith2024} and are distinct for the high- and low-$\alpha$ disks.  While the ability of 2-process models to predict a host of $\alpha$ and Fe-peak elements within 0.05 dex suggests that at this level, our models are capturing the integrated enrichment history of the disk rather than the local star formation history, these residual correlations may still yield information about the overall star formation history of the Milky Way or spatial variations in the interstellar medium due to local nucleosynthesis and mixing.

The potential to discriminate between mass ranges and timescales as we have seen with the \textit{s}-process elements and CCSNe versus SNIa elements suggest an exciting application of this method towards constraining the initial mass function and star formation history of galaxies in conjunction with galactic chemical evolution models such as \citet{Chempy}, \citet{OMEGA}, and \citet{Kobayashi2020} and star-by-star hydrodynamic simulations such as {\sc Aeos} \citep{AeosMethods} and {\sc Inferno} \citep{Inferno2023,Inferno2025}.  By constraining the effects that the initial mass function and star formation history have on a linear model that describes stellar abundances, we can not only use this as a method to understand the chemical histories of individual galaxies, but to also decode the assembly history of our own Galaxy.

Additional samples of high precision data over a wider metallicity range, that cover a wide range of elements, will enable us to probe the detailed nucleosynthetic origins of elements, as well as the history of star formation in the Milky Way.  Similarly, models such as this could be applied to Local Group dwarf galaxies to understand their star formation and enrichment histories and how they differ.

\subsection{The Future of Chemical Tagging and Large Surveys}

In line with previous studies, our work demonstrates the necessity of high-precision chemical abundances that reach below the $2-5\%$ level in order to access the scatter uncoupled to other elements and achieve the high-dimensional chemical space \citep[$C$-space;][]{Freeman2002} necessary to realize chemical tagging.  Higher-precision measurements do not necessarily require higher-resolution spectroscopy and can in principle be achieved with lower resolution spectroscopy with higher signal-to-noise ratio (SNR) \citep[][]{Ting2017}.  For upcoming surveys such as WEAVE, MOONS, and 4MOST, that have R = 4000-6000, a precision of 0.02~dex $(5\%)$ can be achieved for $\sim 30$ elements at a fixed SNR per pixel of 100. However, improving the precision to just 0.01~dex $(2\%)$ limits it to just $\sim 15$ elements.  We note that lower-resolution spectra with the same exposure time and same number of pixels per detector as higher-resolution spectra can achieve the same number of elements measured at similar precision as higher-resolution spectra due to the broader wavelength coverage \citep[][]{Ting2017}.  However, achieving precisions below the 0.01~dex level for enough elements to make chemical tagging feasible would require higher SNR or otherwise high-resolution spectroscopy with broad wavelength coverage.

\subsection{The Dimensionality of Chemical Space}
\subsubsection{$C$-Space in \citetalias{Bedell2018}}
\begin{figure*}
    \centering
    \includegraphics[width=\linewidth,trim={2cm 1cm 2cm 2cm},clip]{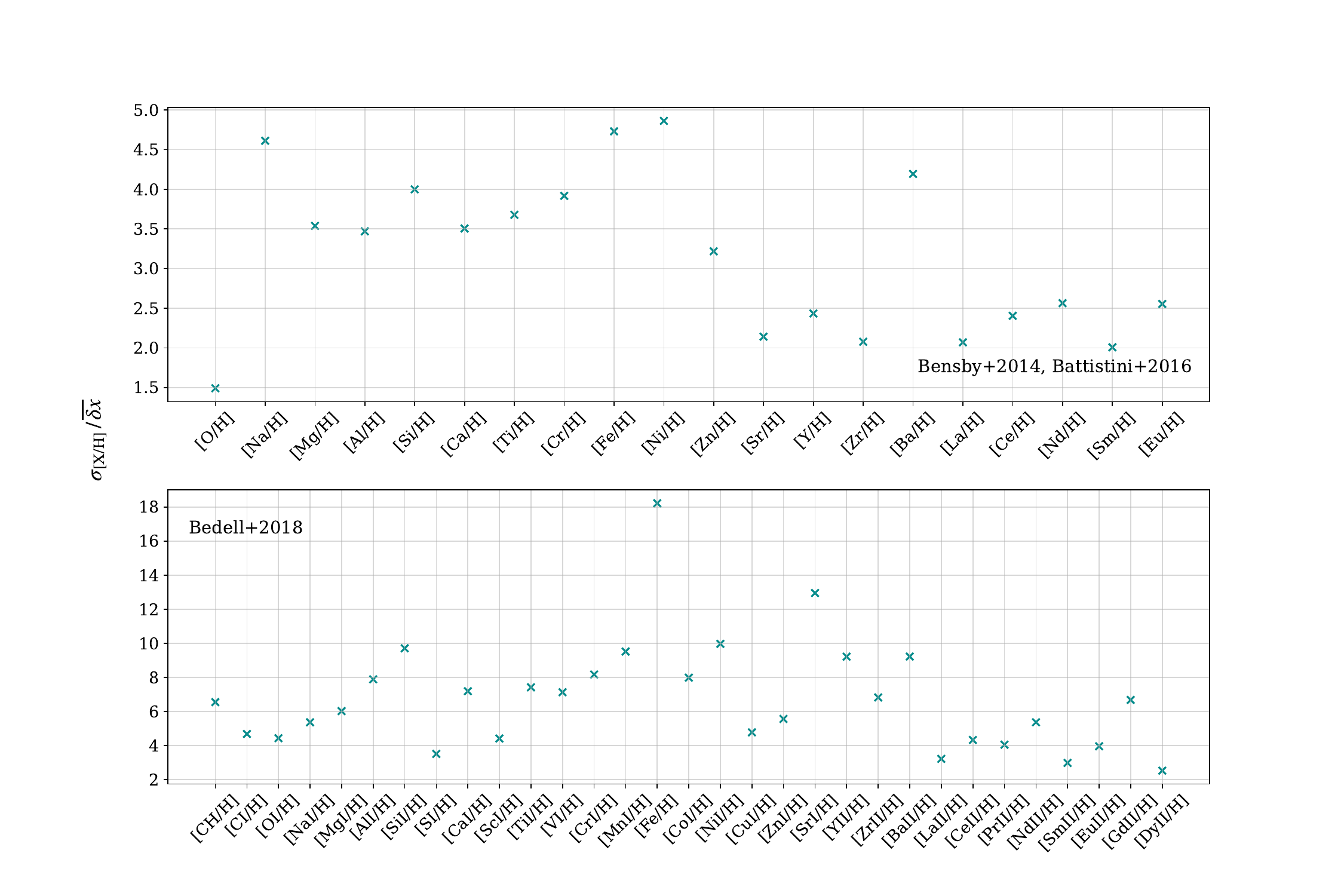}
    \caption{Ratio of the standard deviation of element abundances to the mean uncertainty. Top panel: \citetalias{Bensby2014} and \citetalias{Battistini2016} data. Bottom panel: \citetalias{Bedell2018} data.  Due to the high precision (low uncertainty) of the \citetalias{Bedell2018} data, the mean uncertainty is a factor of four smaller than the standard deviation for nearly all elements, a higher ratio than for any element from \citetalias{Bensby2014} or \citetalias{Battistini2016}.}
    \label{fig:std-err}
\end{figure*}

The analysis we have undertaken allows us to calculate the number of chemical cells, or the dimension of $C$-space, that the high-precision measurements we use can access in practice for the Milky Way. This $C$-space represents the number of unique bins of element abundance patterns, and is a function of the range of the element abundances, the precision of their measurements, and how much independent information they each contain. The amount of independent information in each element can be quantified by the ratio of the standard deviation of the element abundances to the mean uncertainties (Figure \ref{fig:std-err}). If the element abundance measurements were independent, the product of the values in Figure \ref{fig:std-err} would represent this space, which at the precision of \citetalias{Bedell2018} ($\sim0.01$~dex or $2\%$) for 30 elements, is a potential $C$-space of $\sim 10^{23}$; well in excess of the likely number of disk-forming clusters of $\sim10^5 - 10^8$ \citep[see][]{Freeman2002, BH2010}. However, element abundances are highly correlated. Therefore, to calculate the $C$-space in practice, we proceed by first calculating the $C$-space of the elements used as predictors in our full 4-source model. We do this by multiplying the ratios of standard deviation to uncertainties, reported in Figure \ref{fig:std-err} for ([Mg/H], [Y/H], [Ce/H]) only, which results in a $C$-space of 240 bins for these elements alone. We then multiply this value by the residual products. The residual products are the ratios of the intrinsic dispersion of elements predicted by this model as reported in Figure \ref{fig:bedell-full}, to their respective average element abundance uncertainties (i.e. cells of information not captured by the predictor elements), which results in a total of $\sim 2.5\times 10^6$ bins in $C$-space at the precision of \citetalias{Bedell2018} and at solar metallicity. This provides an upper limit on $C$-space in this regime since we do not take into account the remaining residual correlations discussed in Appendix \ref{sec:app-resfull}, but we note that there is substantial additional discriminating power across the disk's range of [Fe/H].  We do not include the $C$-space of [Fe/H] in our calculation at solar metallicity as the $C$-space in for every element depends on the metallicity range, which is arbitrary across the narrow range of [Fe/H] here.  Nevertheless, adopting this approximation highlights the power of high-precision abundance measurements and allows for comparisons between datasets.
  
\subsubsection{$C$-Space in Large Surveys}
To compare the $C$-space achieved by high-precision, high-fidelity datasets to large surveys, we calculate the abundance $C$-space at solar metallicity for both the large GALAH DR4 \citep{Buder2024} and APOGEE DR17 \citep{dr17} surveys. We proceed by calculating the discriminating power in the elements used as predictors in our models, using the survey abundance measurements and reported uncertainties. Similarly to Figure \ref{fig:std-err}, we calculate the ratio of the standard deviation of the element abundances to the mean uncertainty for ([Mg/H], [Ce/H]) in APOGEE and ([Mg/H], [Ce/H], [Y/H]) in GALAH, which sets the majority of the $C$-space at the precision of these surveys. We use the intrinsic scatter calculations for the full 4-source model for GALAH and the 3-source model with [Ce/H] for APOGEE reported in Figures \ref{fig:bedell-full} and \ref{fig:IS_3proc} respectively, to evaluate the contribution of the elements beyond the primary sources to the $C$-space. We do this by multiplying the ratios of the intrinsic scatter to the median uncertainties reported by the surveys for ratios $>$ 1 which thus contribute to the $C$-space. In this way we measure the dimensionality from a set of primary elements and then the remaining contribution from additional elements. We also test the $C$-space across the full disk metallicity range by evaluating the element discriminating power for stars across metallicities $-1.0 <$ [Fe/H] $<$ 0.5, in which case we additionally multiply by the $C$-space of [Fe/H] derived from Figure \ref{fig:std-err}.

APOGEE measures the abundances of 19 elements for $\sim$700,000 stars, including C, N, O, Na, Mg, Al, Si, P, K, Ca, Ti, V, Cr, Mn, Fe, Co, Ni, and Ce with spectra of $R = 22,500$ and $\lambda~=~1510\,-\,1700$ nm. GALAH measures up to 30 elements for $\sim$900,000 stars, including many neutron capture elements, C, N, O, Na, Mg, K, Al, Si, Ca, Ti, Sc, V, Cr, Mn, Co, Ni, Cu, Zn, Rb, Sr, Y, Zr, Mo, Ba, La, Ce, Nd, Ru, Sm, and Eu using spectra with $R=28,000$ and $\lambda~=~471.3-490.3\,{\rm nm}$ / $564.8-587.3\,{\rm nm}$ / $647.8-673.7\,{\rm nm}$ / $758.5-788.7\,{\rm nm}$. Both APOGEE and GALAH report their abundances and abundance uncertainties with respect to Fe. To estimate the C-space for these surveys, we convert the abundance ratios to be with respect to H ([X/H] = [X/Fe] + [Fe/H]), and estimate the uncertainties as the quadrature sum of the reported errors on [Fe/H] and [X/Fe].

At the median SNR of GALAH (SNR=30), the median [X/Fe] and [Fe/H] precisions are $\sigma_{[Fe/H],[X/Fe]}=$~0.06 dex (15 percent). These uncertainties limit the $C$-space to $\sim3$, at solar metallicity for the median survey data (and $\sim$60 across $-1 <$ [Fe/H] $< 0.5$).  For the small red giant subset (2 $<$ $\log g$ $<$ 3) of GALAH stars with SNR $>$ 100, which comprise only 0.3 percent of the survey, the median element abundance precision for the $\sim$30 elements is high, at $\sigma_{[X/Fe]}\sim$~0.02 dex ($\sim$5 percent). However, this does not expand the GALAH $C$-space substantially, as the reported median [Fe/H] uncertainty, even for this highest-quality subset of data is $\sigma_{[Fe/H]}=$~0.05.  Across $-1 <$ [Fe/H] $< 0.5$ for the disk, $C$-space increases to $\sim$200, but at solar metallicity this does not substantially change. A higher [Fe/H] precision for GALAH, of $\sigma_{[X/Fe]}\sim$~0.01~dex can be obtained  \citep[see][]{Manea2024}, which would enable an increase of the GALAH $C$-space. Note, we include N, Zn, Rb and Mo in our calculations (in principle, although in practice these uncertainties are higher than the assumed residuals), which are measured for GALAH but are not part of the B18 analysis, by assuming that these have similar intrinsic scatters to the elements that are evaluated measured in Figure \ref{fig:bedell-full}. In practice, the full $C$-space contribution for GALAH comes from the elements used as predictors in the model, although with higher precision [Fe/H] uncertainties, the other elements would likely contribute.

APOGEE reports fewer elements than GALAH, and only one neutron capture element, but the spectra have a high median SNR = 140. The median abundance precisions of the survey elements are therefore higher than GALAH for many elements, with $\sigma_{[X/Fe]}$$\sim$~0.02 dex, and $\sigma_{[Fe/H]}$$=$~0.01 dex. However, the uncertainties of the elements used as model predictors are most important, in this case (Fe, Mg and Ce). The median reported uncertainties of [Mg/H] and [Ce/H] are $\sigma_{[X/H]}=$ 0.018, 0.08~dex respectively; the uncertainty on [Ce/H] is limiting and this does not decrease substantially with SNR because the precision is limited by line depth (even at SNR $>$~300, $\sigma_{[Ce/Fe]}$$=$~0.06~dex). The APOGEE measurements return a $C$-space of 25 at solar metallicity for the predictors used in the 3-source model with [Ce/H]. We allow the inclusion of N, K, and P, elements measured in APOGEE but not \citetalias{Bedell2018}, by assuming that these have similar intrinsic scatter to elements that are evaluated measured in this analysis. However, of all elements not in the source model only [C/H] and [Mn/H] have uncertainties smaller than the respective intrinsic scatter. The $C$-space for APOGEE increases to $\sim$ 3000 across the full disk metallicity range $-1 <$ [Fe/H] $<$ 0.5, for the median survey data. For highest quality 10\% of the survey with SNR $>$ 350, this $C$ space increases to $\sim$10,000.

We undertake a final test where we evaluate the $C$-space for the $\sim$~60,000 GALAH stars in common with APOGEE, allowing us to adopt the reported [Fe/H] from APOGEE with its small uncertainties and the $\sim$~29 GALAH elements as reported from the optical spectra. In this case, the $C$-space for GALAH across the full disk range of $-1 <$ [Fe/H] $< 0.5$ is $\sim$2000. For the small subset of $\sim$~400 red giant stars with GALAH SNR $>$ 100 across -1 < [Fe/H] $<$ 0.5, this increases to $\sim$200,000. The elements C, Na, Ti, Mn, Co, Ni, Cu have residual scatters in excess of the uncertainties and subsequently also contribute slightly, by a factor of 4, to the total $C$-space.

\subsubsection{Limitations on the Estimates of $C$-Space}
These comparisons highlight that both element coverage and high precision measurements are extremely important for accessing the small amplitudes of independent information in abundances in the Milky Way disk. The neutron capture elements, in addition to the light, $\alpha$, and Fe-peak are extremely valuable for disk analyses to tap into additional overall discriminating power of element abundances in the disk population -- so long as these are measured at precisions of $\lesssim$ 0.02 dex (Manea et al., in prep). An exception to this is that the small fraction of outlier populations, such as Ba-enriched and lithium (Li)-enriched stars are able to be identified at substantially lower precision \citep[e.g.][]{Sayeed2024}.

The small sample of \citetalias{Bedell2018} data reveals the precisions required to achieve a $C$-space approaching the lower limit of what is necessary for chemical tagging in large surveys. However, we emphasize that our $C$-space estimates are optimistic, and an upper limit, as we do not account for residual correlations in the 4-source model, as shown in Figure \ref{fig:res-full}. While we report the $C$-space, we do not claim that the residual information expressed in Figure \ref{fig:bedell-full} is connected to birth cluster sites and it may not offer discriminatory power to separate them. Instead, these conditional abundance residuals may reflect global changes in star formation across the disk \citep[][]{Griffith2024}, as well as stochastic processes. Their amplitudes are also similar to the measurements of intrinsic scatters in open cluster birth sites which are typically 0.01-0.03 dex \citep[e.g.][]{Sinha2024, Ness2022, Cheng2021, Poovelil2020, Liu2019, Bovy2016}. Systematic abundance variations across evolutionary state can also inflate this measurement, and these trends can be substantial \citep{Kos2025, Jofre2019}. Nevertheless, maximizing this space via high precision abundances enables the disk to be studied in finer detail and the $C$-space amplitude is an important property of a survey.

\section{Conclusions} \label{sec:conc}
In this work, we demonstrate that we are able to predict 30 $\alpha$, Fe-peak, and neutron capture abundances of solar neighborhood stars using only four element abundances representing four nucleosynthetic sources along with $T_{\rm eff}$ and log$g$.  For the first time, we show that the \textit{s}-process component from AGB stars must be split into two, with a contribution from the first and second peak \textit{s}-process elements in the source model, to  obtain additional predictive power of other neutron capture elements, compared to any single \textit{s}-process source. A single representative element abundance from each of these components is sufficient to predict all other \textit{s}-process abundances to on average 2.5\%.

We build linear models of 30 element abundances in \citet{Bedell2018} and 19 element abundances across \citet{Bensby2014} and \citet{Battistini2016} for every star in each sample using only $T_{\rm eff}$, log$g$, [Fe/H], and [Mg/H] in our fiducial model.  To test the robustness of using Mg and Fe as representative $\alpha$ and Fe-peak elements, we test substituting [Si/H] in for [Mg/H] in the model, as well as [Ni/H] and [Mn/H] for [Fe/H], but find that the fiducial model is robust to these changes, with only some Fe-peak elements showing significant improvement when [Fe/H] is replaced.  We demonstrate that while our fiducial model is able to predict most $\alpha$ and Fe-peak element abundances within $0.02-0.03$~dex, or $5-7\%$, and neutron capture elements abundances to $0.03-0.045$~dex, or $7-11\%$, there remain correlations among the residuals of the model, primarily among the \textit{s}-process elements, and between the \textit{s}-process elements and some $\alpha$ and Fe-peak elements, suggesting that there are processes that are unaccounted for in our modeling.

To explore this, we develop a 3-source model that, in addition to the 2-source fiducial model, includes a neutron capture element in the model.  We find that first peak \textit{s}-process element abundances are able to predict other first peak \textit{s}-process element abundances up to 0.03~dex better than the fiducial model, and similarly that second peak \textit{s}-process element abundances are able to predict other second peak \textit{s}-process element abundances up to 0.03~dex better than the fiducial model.  This is a discriminator of the mass and metallicity of AGB stars that contributed to this enrichment.  We also find that the second peak \textit{s}-process elements are able to improve the prediction for a handful of $\alpha$ and Fe-peak elements, for which the residuals were strongly correlated, and hypothesize that this is related to the timescale overlap between intermediate mass AGB stars and CCSNe, which conversely would result in an anti-correlation with elements primarily produced by SNIa.  We also explore if the accuracy of the predictions of \textit{r}-process elements can be improved by including an \textit{r}-process element, and find that \textit{r}-process elements do not improve the prediction on other \textit{r}-process elements over the fiducial model, suggesting that there are likely multiple sources of \textit{r}-process that produce these elements in different ratios, and possibly with significant stochasticity.

We conclude our work by developing a full, 4-source, model that is able to improve predictions for \textit{s}-process elements by $0.02-0.03$~dex over the fiducial model, and for some $\alpha$ and Fe-peak elements by up to 0.01~dex.  This model represents 4 known nucleosynthetic processes with which all other element abundances in our sample are degenerate at the $2-5\%$ level for most elements, greatly limiting the dimensionality of chemical space.  We show this for a small sample of high-fidelity data with precision and element coverage higher than most large surveys. This strongly motivates the need for high SNR ($> 100$) spectroscopy in future disk surveys to measure $>$ 30 element abundances at precisions of sub $2-5\%$.

Large ($> 10^5-10^6$ star) future spectroscopic survey precision goals can be informed on a per-element basis by our analysis and results. In the regime of such highly correlated abundances, high precision and wide element coverage are key to uncover the detailed history of the temporal and spatial chemical evolution of the Milky Way, to study subtle binary or planetary interactions across populations via anomalously correlated residuals, and to pursue chemical tagging. A complement of  small high-fidelity samples such as those we have utilized in our analysis (79 and 593 stars, respectively) sets the stage for the next generation of surveys and demonstrate the power of high-precision measurements to tap into the nucleosynthetic history of the Milky Way. Additional small high-precision datasets across a wider metallicity range, and in particular those measured using differential abundance techniques, would be extremely valuable to extend this analysis to a wider metallicity or spatial realm.

Our results provide strong constraints on the disk's chemical evolution history. However, it remains to be understood \textit{why} the element abundances of the disk are so rigidly coupled. The measured correlation structure and residual amplitudes can be incorporated into chemical evolution models to learn the parameters that underlie these conditions, and using differential abundance measurements circumvents the need to match absolute amplitudes. While the origin of the highly-correlated disk remains to be understood, it may be an outcome of smoothly varying star formation history over time,  such as represented by a chemical equilibrium process \citep{Johnson2024}.

We conclude that pursuing the correlations in conditional abundance residual space is a promising direction for understanding the chemodynamical evolution of the Milky Way disk. Although the rigid coupling of the element abundances of the Milky Way disk is prohibitive for chemical tagging, abundances measured with precision of a few percent capture critical information to differentiate birth sites over spatial scales that may well resolve bulk birth sites or origins, if not individual clusters.

\section*{Acknowledgments}
The authors thank Amanda Karakas for insights on the nature of \textit{s}- and \textit{r}-process element production, and Gary Da Costa for a title idea forged in the fires of Mount Clever itself -- you have our gratitude, our admiration... and our axe.

JM acknowledges support from the National Science Foundation Graduate Research Fellowship under grant DGE-2036197.

\bibliographystyle{yahapj}
\bibliography{refs}

\appendix
\section{Models} \label{sec:app-model}
\begin{longtable}{llllllll}
\caption{Average coefficients and standard deviations over single star models.}
\label{tab:models}\\
\hline
\textbf{Model} & $\mathbf{T_{\rm eff}}$ & $\mathbf{\log g}$ & \textbf{[Fe/H]} & \textbf{[Mg/H]} & \textbf{[Y/H]} & \textbf{[Ce/H]} & \textbf{Intercept}\\
               & $10^{-5}$ [1/K] & $10^{-1}$ & $10^{-1}$ [dex] & $10^{-1}$ [dex] & $10^{-1}$ [dex] & $10^{-1}$ [dex] & [dex]\\
\hline \hline
\textbf{\citetalias{Bensby2014} + \citetalias{Battistini2016} Fiducial} \\
\hline
\ [O/H] & $-10.22 \pm 0.07$ & $1.29 \pm 0.01$ & $-2.67 \pm 0.02$ & $9.68 \pm 0.03$ & - & - & $0.09 \pm 0.06$\\
\ [Na/H] & $4.06 \pm 0.02$ & $-0.45 \pm 0.0$ & $7.39 \pm 0.01$ & $3.37 \pm 0.01$ & - & - & $-0.44 \pm 0.02$\\
\ [Mg/H] & $0.0 \pm 0.0$ & $0.0 \pm 0.0$ & $0.0 \pm 0.0$ & $10.0 \pm 0.0$ & - & - & $0.0 \pm 0.0$\\
\ [Al/H] & $-11.33 \pm 0.03$ & $0.21 \pm 0.0$ & $3.01 \pm 0.01$ & $7.85 \pm 0.02$ & - & - & $5.81 \pm 0.02$\\
\ [Si/H] & $-3.97 \pm 0.02$ & $0.28 \pm 0.0$ & $4.65 \pm 0.01$ & $5.29 \pm 0.01$ & - & - & $1.21 \pm 0.02$\\
\ [Ca/H] & $1.65 \pm 0.02$ & $-0.33 \pm 0.0$ & $5.61 \pm 0.01$ & $3.23 \pm 0.01$ & - & - & $0.5 \pm 0.02$\\
\ [Ti/H] & $-7.7 \pm 0.03$ & $0.76 \pm 0.0$ & $4.11 \pm 0.01$ & $5.08 \pm 0.01$ & - & - & $1.4 \pm 0.03$\\
\ [Cr/H] & $-6.28 \pm 0.02$ & $0.48 \pm 0.0$ & $9.87 \pm 0.01$ & $0.09 \pm 0.01$ & - & - & $1.62 \pm 0.02$\\
\ [Ni/H] & $-5.01 \pm 0.01$ & $0.34 \pm 0.0$ & $8.81 \pm 0.0$ & $2.16 \pm 0.01$ & - & - & $1.3 \pm 0.01$\\
\ [Zn/H] & $-11.32 \pm 0.04$ & $0.79 \pm 0.01$ & $4.06 \pm 0.01$ & $7.2 \pm 0.02$ & - & - & $3.11 \pm 0.03$\\
\ [Sr/H] & $4.75 \pm 0.47$ & $1.4 \pm 0.05$ & $6.56 \pm 0.12$ & $1.88 \pm 0.16$ & - & - & $-9.82 \pm 0.25$\\
\ [Y/H] & $-9.57 \pm 0.06$ & $0.69 \pm 0.01$ & $10.89 \pm 0.02$ & $-1.84 \pm 0.03$ & - & - & $2.49 \pm 0.06$\\
\ [Zr/H] & $7.26 \pm 0.16$ & $0.07 \pm 0.02$ & $6.88 \pm 0.06$ & $-0.59 \pm 0.08$ & - & - & $-4.17 \pm 0.14$\\
\ [Ba/H] & $12.48 \pm 0.07$ & $-1.04 \pm 0.01$ & $13.6 \pm 0.03$ & $-5.55 \pm 0.04$ & - & - & $-2.44 \pm 0.07$\\
\ [La/H] & $10.19 \pm 0.16$ & $0.61 \pm 0.02$ & $8.31 \pm 0.05$ & $-2.06 \pm 0.07$ & - & - & $-8.78 \pm 0.09$\\
\ [Ce/H] & $12.41 \pm 0.11$ & $0.46 \pm 0.01$ & $11.72 \pm 0.04$ & $-5.21 \pm 0.05$ & - & - & $-9.23 \pm 0.08$\\
\ [Nd/H] & $-1.01 \pm 0.09$ & $1.19 \pm 0.01$ & $8.39 \pm 0.03$ & $-1.99 \pm 0.04$ & - & - & $-4.59 \pm 0.08$\\
\ [Sm/H] & $9.52 \pm 0.18$ & $0.28 \pm 0.02$ & $6.39 \pm 0.05$ & $-0.48 \pm 0.07$ & - & - & $-6.0 \pm 0.12$\\
\ [Eu/H] & $7.04 \pm 0.08$ & $0.64 \pm 0.01$ & $2.93 \pm 0.03$ & $3.82 \pm 0.04$ & - & - & $-6.38 \pm 0.07$\\
\hline
\textbf{\citetalias{Bedell2018} Fiducial} \\
\hline
\ [CH/H] & $4.49 \pm 0.74$ & $-2.08 \pm 0.06$ & $7.10 \pm 0.11$ & $2.48 \pm 0.12$ & - & - & $6.18 \pm 0.53$\\
\ [C \RomanNumeralCaps{1}/H] & $2.75 \pm 0.92$ & $-2.77 \pm 0.07$ & $0.32 \pm 0.12$ & $6.95 \pm 0.13$ & - & - & $10.28 \pm 0.6$\\
\ [O \RomanNumeralCaps{1}/H] & $-0.33 \pm 0.52$ & $-0.78 \pm 0.05$ & $-0.12 \pm 0.09$ & $7.09 \pm 0.1$ & - & - & $3.74 \pm 0.38$\\
\ [Na \RomanNumeralCaps{1}/H] & $1.33 \pm 0.74$ & $0.05 \pm 0.08$ & $5.22 \pm 0.09$ & $5.71 \pm 0.12$ & - & - & $-1.26 \pm 0.53$\\
\ [Mg \RomanNumeralCaps{1}/H] & $-0.0 \pm 0.0$ & $0.0 \pm 0.0$ & $0.0 \pm 0.0$ & $10.0 \pm 0.0$ & - & - & $0.0 \pm 0.0$\\
\ [Al \RomanNumeralCaps{1}/H] & $-6.62 \pm 0.31$ & $-0.76 \pm 0.03$ & $0.14 \pm 0.06$ & $10.26 \pm 0.07$ & - & - & $7.15 \pm 0.27$\\
\ [Si \RomanNumeralCaps{1}/H] & $3.08 \pm 0.15$ & $-0.15 \pm 0.02$ & $4.04 \pm 0.02$ & $5.6 \pm 0.03$ & - & - & $-1.2 \pm 0.13$\\
\ [S \RomanNumeralCaps{1}/H] & $-12.43 \pm 0.74$ & $-1.42 \pm 0.09$ & $4.58 \pm 0.13$ & $3.83 \pm 0.15$ & - & - & $13.35 \pm 0.58$\\
\ [Ca \RomanNumeralCaps{1}/H] & $2.17 \pm 0.29$ & $1.13 \pm 0.04$ & $5.71 \pm 0.06$ & $3.86 \pm 0.06$ & - & - & $-6.09 \pm 0.25$\\
\ [Sc \RomanNumeralCaps{1}/H] & $10.15 \pm 0.37$ & $1.18 \pm 0.04$ & $3.04 \pm 0.07$ & $7.26 \pm 0.08$ & - & - & $-11.14 \pm 0.33$\\
\ [Ti \RomanNumeralCaps{1}/H] & $-2.17 \pm 0.32$ & $1.68 \pm 0.04$ & $1.64 \pm 0.06$ & $8.24 \pm 0.07$ & - & - & $-6.09 \pm 0.3$\\
\ [V \RomanNumeralCaps{1}/H] & $-1.18 \pm 0.21$ & $1.55 \pm 0.03$ & $4.42 \pm 0.04$ & $6.48 \pm 0.05$ & - & - & $-6.2 \pm 0.22$\\
\ [Cr \RomanNumeralCaps{1}/H] & $-5.32 \pm 0.13$ & $0.88 \pm 0.01$ & $9.98 \pm 0.02$ & $0.82 \pm 0.02$ & - & - & $-0.79 \pm 0.12$\\
\ [Mn \RomanNumeralCaps{1}/H] & $-12.16 \pm 0.47$ & $-0.2 \pm 0.05$ & $13.69 \pm 0.08$ & $-1.84 \pm 0.09$ & - & - & $7.75 \pm 0.37$\\
\ [Co \RomanNumeralCaps{1}/H] & $2.83 \pm 0.36$ & $0.28 \pm 0.04$ & $3.73 \pm 0.05$ & $6.77 \pm 0.06$ & - & - & $-3.08 \pm 0.28$\\
\ [Ni \RomanNumeralCaps{1}/H] & $-4.97 \pm 0.38$ & $-0.5 \pm 0.04$ & $8.57 \pm 0.06$ & $2.69 \pm 0.07$ & - & - & $4.87 \pm 0.31$\\
\ [Cu \RomanNumeralCaps{1}/H] & $-7.21 \pm 0.55$ & $-1.83 \pm 0.06$ & $4.48 \pm 0.09$ & $6.89 \pm 0.11$ & - & - & $11.99 \pm 0.43$\\
\ [Zn \RomanNumeralCaps{1}/H] & $-9.55 \pm 0.57$ & $-0.34 \pm 0.05$ & $2.73 \pm 0.07$ & $8.9 \pm 0.08$ & - & - & $6.75 \pm 0.41$\\
\ [Sr \RomanNumeralCaps{1}/H] & $-2.55 \pm 0.84$ & $8.66 \pm 0.12$ & $9.64 \pm 0.16$ & $4.04 \pm 0.22$ & - & - & $-36.67 \pm 0.78$\\
\ [Y \RomanNumeralCaps{2}/H] & $-5.57 \pm 0.73$ & $7.94 \pm 0.11$ & $10.03 \pm 0.13$ & $2.95 \pm 0.17$ & - & - & $-31.86 \pm 0.73$\\
\ [Zr \RomanNumeralCaps{2}/H] & $-1.91 \pm 0.75$ & $7.45 \pm 0.11$ & $7.61 \pm 0.13$ & $3.63 \pm 0.17$ & - & - & $-31.66 \pm 0.73$\\
\ [Ba \RomanNumeralCaps{2}/H] & $17.23 \pm 1.09$ & $7.05 \pm 0.12$ & $9.91 \pm 0.16$ & $0.84 \pm 0.18$ & - & - & $-40.68 \pm 1.02$\\
\ [La \RomanNumeralCaps{2}/H] & $7.7 \pm 1.05$ & $6.62 \pm 0.13$ & $4.0 \pm 0.17$ & $4.17 \pm 0.19$ & - & - & $-33.4 \pm 0.95$\\
\ [Ce \RomanNumeralCaps{2}/H] & $13.01 \pm 0.94$ & $5.4 \pm 0.11$ & $5.95 \pm 0.15$ & $2.34 \pm 0.16$ & - & - & $-30.94 \pm 0.88$\\
\ [Pr \RomanNumeralCaps{2}/H] & $6.33 \pm 0.86$ & $3.62 \pm 0.11$ & $1.93 \pm 0.19$ & $5.0 \pm 0.18$ & - & - & $-18.63 \pm 0.81$\\
\ [Nd \RomanNumeralCaps{2}/H] & $17.03 \pm 0.98$ & $5.08 \pm 0.11$ & $4.08 \pm 0.15$ & $3.78 \pm 0.15$ & - & - & $-31.7 \pm 0.86$\\
\ [Sm \RomanNumeralCaps{2}/H] & $8.21 \pm 0.69$ & $3.07 \pm 0.08$ & $2.58 \pm 0.1$ & $5.82 \pm 0.11$ & - & - & $-18.05 \pm 0.6$\\
\ [Eu \RomanNumeralCaps{2}/H] & $23.26 \pm 0.55$ & $2.09 \pm 0.07$ & $2.06 \pm 0.1$ & $6.25 \pm 0.11$ & - & - & $-22.11 \pm 0.43$\\
\ [Gd \RomanNumeralCaps{2}/H] & $6.23 \pm 0.92$ & $2.47 \pm 0.09$ & $2.54 \pm 0.14$ & $6.75 \pm 0.17$ & - & - & $-14.25 \pm 0.7$\\
\ [Dy \RomanNumeralCaps{2}/H] & $7.18 \pm 1.09$ & $2.79 \pm 0.11$ & $2.0 \pm 0.19$ & $5.47 \pm 0.22$ & - & - & $-16.0 \pm 0.95$\\
\hline
\textbf{\citetalias{Bedell2018} 4-source} \\
\hline
\ [CH/H] & $9.14 \pm 0.72$ & $-1.23 \pm 0.11$ & $7.9 \pm 0.17$ & $2.89 \pm 0.13$ & $1.06 \pm 0.14$ & $-3.12 \pm 0.11$ & $-0.12 \pm 0.64$\\
\ [C \RomanNumeralCaps{1}/H] & $7.66 \pm 0.92$ & $-1.67 \pm 0.13$ & $1.4 \pm 0.19$ & $7.46 \pm 0.14$ & $0.92 \pm 0.19$ & $-3.38 \pm 0.15$ & $2.76 \pm 0.74$\\
\ [O \RomanNumeralCaps{1}/H] & $-2.9 \pm 0.56$ & $-0.95 \pm 0.07$ & $-0.21 \pm 0.09$ & $6.98 \pm 0.11$ & $-0.87 \pm 0.06$ & $1.61 \pm 0.08$ & $5.92 \pm 0.46$\\
\ [Na \RomanNumeralCaps{1}/H] & $10.81 \pm 0.63$ & $1.11 \pm 0.12$ & $6.02 \pm 0.15$ & $6.3 \pm 0.1$ & $2.8 \pm 0.17$ & $-6.08 \pm 0.11$ & $-11.15 \pm 0.65$\\
\ [Mg \RomanNumeralCaps{1}/H] & $0.0 \pm 0.0$ & $0.0 \pm 0.0$ & $0.0 \pm 0.0$ & $10.0 \pm 0.0$ & $0.0 \pm 0.0$ & $0.0 \pm 0.0$ & $0.0 \pm 0.0$\\
\ [Al \RomanNumeralCaps{1}/H] & $-6.31 \pm 0.3$ & $0.13 \pm 0.05$ & $1.22 \pm 0.06$ & $10.61 \pm 0.07$ & $-0.75 \pm 0.03$ & $-0.55 \pm 0.04$ & $3.05 \pm 0.31$\\
\ [Si \RomanNumeralCaps{1}/H] & $4.31 \pm 0.21$ & $-0.08 \pm 0.06$ & $4.05 \pm 0.09$ & $5.65 \pm 0.03$ & $0.44 \pm 0.11$ & $-0.76 \pm 0.06$ & $-2.16 \pm 0.28$\\
\ [S \RomanNumeralCaps{1}/H] & $-10.69 \pm 0.77$ & $-1.05 \pm 0.1$ & $4.94 \pm 0.14$ & $4.01 \pm 0.16$ & $0.34 \pm 0.07$ & $-1.18 \pm 0.11$ & $10.77 \pm 0.62$\\
\ [Ca \RomanNumeralCaps{1}/H] & $-1.47 \pm 0.26$ & $0.4 \pm 0.07$ & $5.01 \pm 0.11$ & $3.51 \pm 0.06$ & $-0.77 \pm 0.12$ & $2.47 \pm 0.07$ & $-0.91 \pm 0.3$\\
\ [Sc \RomanNumeralCaps{1}/H] & $8.42 \pm 0.41$ & $1.18 \pm 0.07$ & $3.12 \pm 0.1$ & $7.22 \pm 0.09$ & $-0.69 \pm 0.08$ & $1.03 \pm 0.07$ & $-10.15 \pm 0.46$\\
\ [Ti \RomanNumeralCaps{1}/H] & $-6.48 \pm 0.3$ & $1.08 \pm 0.08$ & $1.12 \pm 0.13$ & $7.92 \pm 0.08$ & $-1.15 \pm 0.15$ & $2.81 \pm 0.08$ & $-1.06 \pm 0.38$\\
\ [V \RomanNumeralCaps{1}/H] & $-2.87 \pm 0.25$ & $1.52 \pm 0.07$ & $4.47 \pm 0.09$ & $6.43 \pm 0.06$ & $-0.65 \pm 0.11$ & $1.02 \pm 0.07$ & $-5.12 \pm 0.36$\\
\ [Cr \RomanNumeralCaps{1}/H] & $-6.58 \pm 0.14$ & $0.68 \pm 0.03$ & $9.8 \pm 0.04$ & $0.72 \pm 0.03$ & $-0.32 \pm 0.05$ & $0.83 \pm 0.03$ & $0.77 \pm 0.15$\\
\ [Mn \RomanNumeralCaps{1}/H] & $-5.28 \pm 0.31$ & $0.94 \pm 0.06$ & $14.73 \pm 0.1$ & $-1.27 \pm 0.08$ & $1.67 \pm 0.07$ & $-4.57 \pm 0.05$ & $-1.08 \pm 0.36$\\
\ [Co \RomanNumeralCaps{1}/H] & $6.46 \pm 0.36$ & $1.04 \pm 0.07$ & $4.47 \pm 0.08$ & $7.14 \pm 0.06$ & $0.73 \pm 0.06$ & $-2.48 \pm 0.06$ & $-8.42 \pm 0.45$\\
\ [Ni \RomanNumeralCaps{1}/H] & $1.16 \pm 0.27$ & $0.56 \pm 0.07$ & $9.55 \pm 0.11$ & $3.22 \pm 0.05$ & $1.44 \pm 0.12$ & $-4.09 \pm 0.07$ & $-3.19 \pm 0.32$\\
\ [Cu \RomanNumeralCaps{1}/H] & $1.02 \pm 0.41$ & $-0.78 \pm 0.08$ & $5.34 \pm 0.13$ & $7.45 \pm 0.11$ & $2.31 \pm 0.1$ & $-5.34 \pm 0.08$ & $2.82 \pm 0.47$\\
\ [Zn \RomanNumeralCaps{1}/H] & $-6.61 \pm 0.58$ & $-0.14 \pm 0.07$ & $2.83 \pm 0.09$ & $9.04 \pm 0.09$ & $0.99 \pm 0.05$ & $-1.84 \pm 0.09$ & $4.22 \pm 0.51$\\
\ [Sr \RomanNumeralCaps{1}/H] & $5.71 \pm 0.32$ & $-0.58 \pm 0.09$ & $-2.12 \pm 0.14$ & $0.64 \pm 0.05$ & $12.35 \pm 0.18$ & $-1.06 \pm 0.09$ & $-0.59 \pm 0.37$\\
\ [Y \RomanNumeralCaps{2}/H] & $0.0 \pm 0.0$ & $0.0 \pm 0.0$ & $0.0 \pm 0.0$ & $0.0 \pm 0.0$ & $10.0 \pm 0.0$ & $0.0 \pm 0.0$ & $0.0 \pm 0.0$\\
\ [Zr \RomanNumeralCaps{2}/H] & $0.42 \pm 0.24$ & $-0.06 \pm 0.03$ & $-1.71 \pm 0.04$ & $0.78 \pm 0.05$ & $8.26 \pm 0.02$ & $1.75 \pm 0.05$ & $0.05 \pm 0.22$\\
\ [Ba \RomanNumeralCaps{2}/H] & $3.78 \pm 0.47$ & $0.83 \pm 0.06$ & $2.98 \pm 0.11$ & $-1.82 \pm 0.11$ & $0.62 \pm 0.05$ & $10.6 \pm 0.08$ & $-5.89 \pm 0.47$\\
\ [La \RomanNumeralCaps{2}/H] & $-1.89 \pm 0.68$ & $0.67 \pm 0.12$ & $-2.81 \pm 0.17$ & $1.69 \pm 0.1$ & $1.93 \pm 0.2$ & $8.19 \pm 0.15$ & $-1.91 \pm 0.64$\\
\ [Ce \RomanNumeralCaps{2}/H] & $-0.0 \pm 0.0$ & $-0.0 \pm 0.0$ & $-0.0 \pm 0.0$ & $0.0 \pm 0.0$ & $0.0 \pm 0.0$ & $10.0 \pm 0.0$ & $0.0 \pm 0.0$\\
\ [Pr \RomanNumeralCaps{2}/H] & $-0.87 \pm 0.88$ & $-0.03 \pm 0.15$ & $-2.17 \pm 0.25$ & $3.45 \pm 0.17$ & $0.64 \pm 0.23$ & $5.8 \pm 0.19$ & $1.38 \pm 0.78$\\
\ [Nd \RomanNumeralCaps{2}/H] & $2.6 \pm 0.34$ & $-0.06 \pm 0.06$ & $-1.47 \pm 0.06$ & $1.52 \pm 0.07$ & $-0.83 \pm 0.05$ & $10.73 \pm 0.06$ & $-1.15 \pm 0.35$\\
\ [Sm \RomanNumeralCaps{2}/H] & $-1.14 \pm 0.43$ & $-0.01 \pm 0.07$ & $-0.71 \pm 0.1$ & $4.45 \pm 0.08$ & $-0.78 \pm 0.11$ & $6.85 \pm 0.09$ & $0.65 \pm 0.42$\\
\ [Eu \RomanNumeralCaps{2}/H] & $17.55 \pm 0.71$ & $0.09 \pm 0.19$ & $-0.09 \pm 0.28$ & $5.37 \pm 0.13$ & $-0.36 \pm 0.34$ & $4.23 \pm 0.2$ & $-10.17 \pm 0.78$\\
\ [Gd \RomanNumeralCaps{2}/H] & $-2.13 \pm 0.77$ & $-0.45 \pm 0.16$ & $-0.6 \pm 0.26$ & $5.47 \pm 0.14$ & $-0.54 \pm 0.3$ & $6.19 \pm 0.19$ & $3.18 \pm 0.71$\\
\ [Dy \RomanNumeralCaps{2}/H] & $-5.65 \pm 0.82$ & $-1.06 \pm 0.1$ & $-2.05 \pm 0.16$ & $3.74 \pm 0.17$ & $-1.44 \pm 0.14$ & $9.25 \pm 0.14$ & $8.01 \pm 0.73$

\end{longtable}

\section{Residual Noise} \label{sec:app-resnoise}

\begin{figure*}[h!]
    \centering
    \includegraphics[width=\linewidth,trim={7cm 4cm 6cm 6cm},clip]{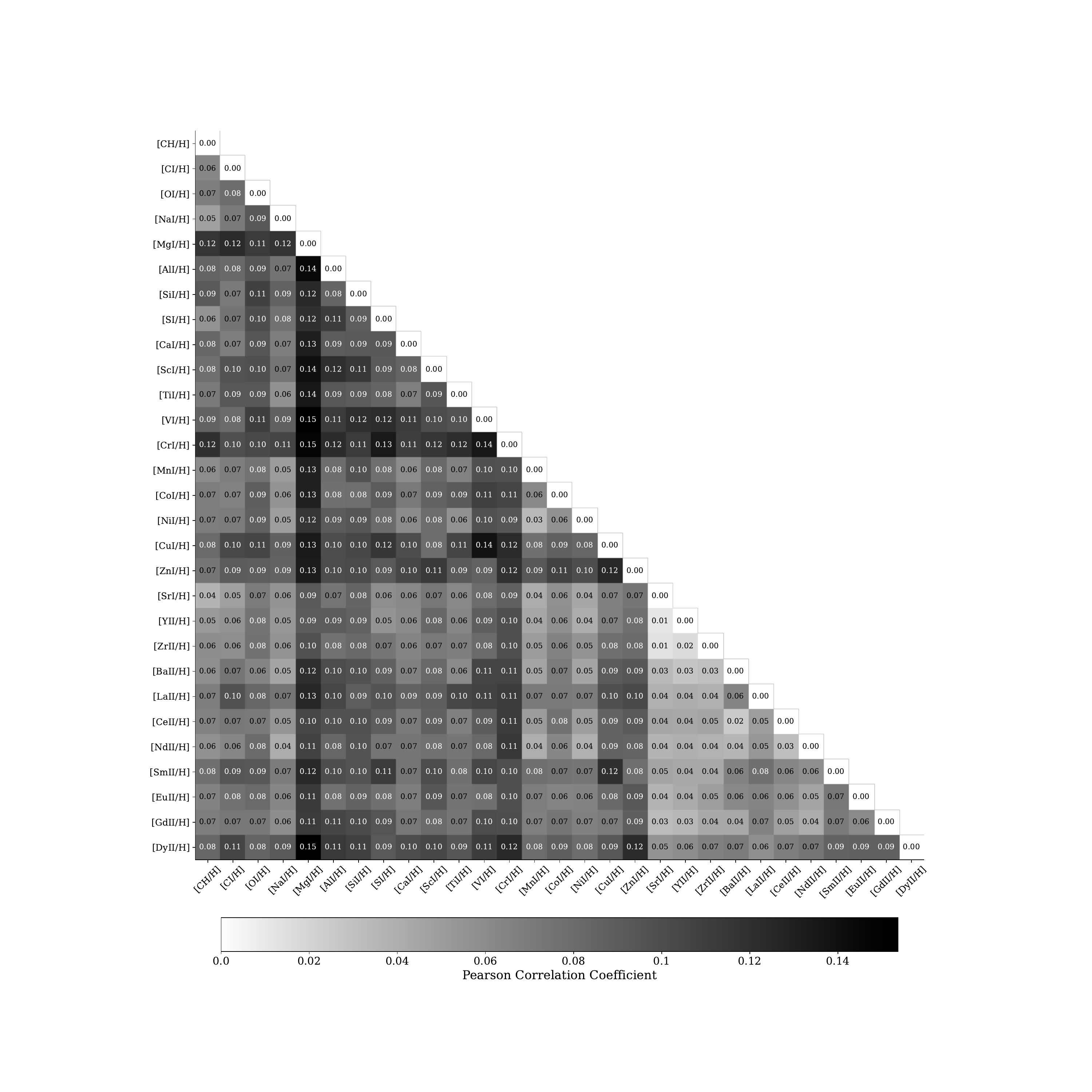}
    \caption{Standard deviation of Pearson correlation coefficients from noise perturbed data.}
    \label{fig:res-noise}
\end{figure*}

We quantify the noise in the correlations of the conditional abundance residuals by perturbing the data by random gaussian noise with standard deviation equal to the uncertainty on the data and recalculating the Pearson correlation coefficient.  We perform this test 50 times and calculate the standard deviation over all correlation coefficients.  The noise level is quantified in Figure \ref{fig:res-noise}.

\section{Residuals from a 4-source model} \label{sec:app-resfull}

\begin{figure*}
    \centering
    \includegraphics[width=0.985\linewidth,trim={7cm 5.2cm 6cm 6cm},clip]{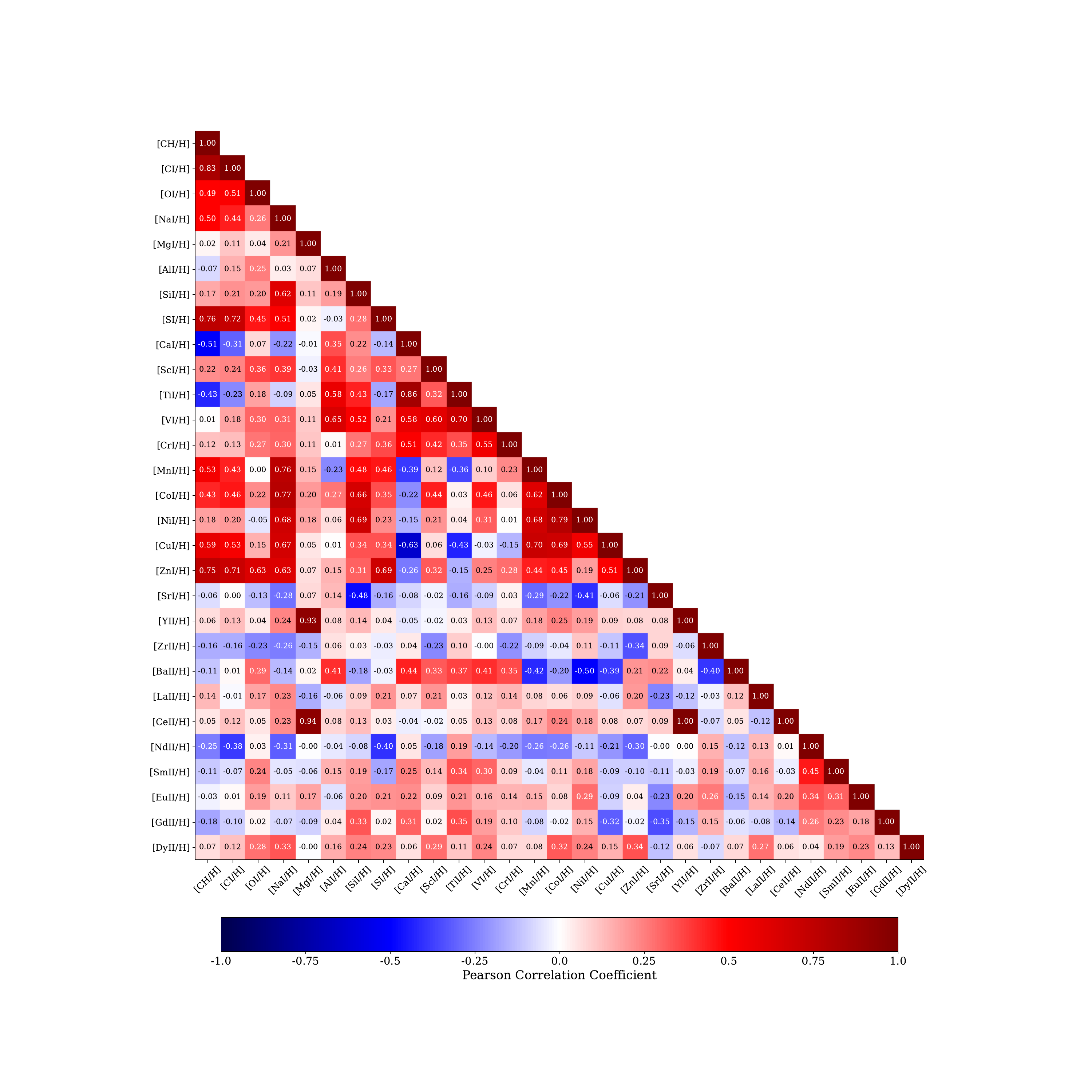}
    \caption{Pearson correlation coefficients of the [X/H] residuals from the 4-source model with every other [X/H] abundance residual.}
    \label{fig:res-full}
\end{figure*}

We demonstrate here that the inclusion of element abundances representative of the first and second peak \textit{s}-process elements in the 4-source model is capable of reducing the strong correlations among neutron capture element residuals that remained in the fiducial model.  Figure \ref{fig:res-full} shows that strong (anti-)correlations between the second peak \textit{s}-process elements and some of the $\alpha$ and Fe-peak elements have been reduced.  Notably, intermediate to strong correlations still remain among $\alpha$ and Fe-peak elements, suggesting either nucleosynthetic sources or physical properties that remain unaccounted for.

\end{document}